\documentclass[%
reprint,
superscriptaddress,
amsmath,amssymb,
aip,
]{revtex4-1}

\usepackage{graphicx}% Include figure files
\usepackage{dcolumn}% Align table columns on decimal point
\usepackage{bm}% bold math
\usepackage{overpic}% super impose on the picture
\usepackage{pict2e}
\usepackage{tikz}

\usepackage{natbib}
\usepackage[nottoc,numbib]{tocbibind}
\usepackage[colorlinks]{hyperref}    % better urls in bibliography
\usepackage{amsmath}
\usepackage{gensymb}

\begin{document}

\preprint{AIP/123-QED}

\title{THz-Pump and X-Ray-Probe Sources Based on an Electron Linac}
%\\with Forced Linebreak\footnote{Error!}}% Force line breaks with \\
%\thanks{Footnote to title of article.}

\author{Sadiq \surname{Setiniyaz}}
\email{sadik82@gmail.com}
\affiliation{Korea Atomic Energy Research Institute, 1045 Daedeok-Daero, Yuseong-gu, Daejeon 34057, Republic of Korea}
%\thanks{Fax: +82-42-866-6150}

\author{Seong Hee \surname{Park}}
\email{shpark7@korea.ac.kr}
\affiliation{Korea University, 2511 Sejong-ro, Sejong 30019, Republic of Korea}

\author{Hyun Woo \surname{Kim}}
\affiliation{Korea Atomic Energy Research Institute, 1045 Daedeok-Daero, Yuseong-gu, Daejeon 34057, Republic of Korea}
\affiliation{University of Science and Technology, 217, Gajeong-ro, Yuseong-gu, Daejeon 34113, Republic of Korea}

\author{Nikolay A. \surname{Vinokurov}}
\affiliation{Budker Institute of Nuclear Physics, Siberian Branch of Russian Academy of Sciences, 11 Lavrentyev Prospect, Novosibirsk 630090, Russia}

\author{Kyu-Ha \surname{Jang}}
\affiliation{Korea Atomic Energy Research Institute, 1045 Daedeok-Daero, Yuseong-gu, Daejeon 34057, Republic of Korea}

\author{Kitae \surname{Lee}}
\affiliation{Korea Atomic Energy Research Institute, 1045 Daedeok-Daero, Yuseong-gu, Daejeon 34057, Republic of Korea}

\author{In Hyung \surname{Baek}}
\affiliation{Korea Atomic Energy Research Institute, 1045 Daedeok-Daero, Yuseong-gu, Daejeon 34057, Republic of Korea}

\author{Young Uk \surname{Jeong}}
\affiliation{Korea Atomic Energy Research Institute, 1045 Daedeok-Daero, Yuseong-gu, Daejeon 34057, Republic of Korea}

\date{\today}% It is always \today, today,
%  but any date may be explicitly specified

\begin{abstract}	
	We describe a compact THz-pump and X-ray-probe beamline, based on an electron linac, for ultrafast time-resolved diffraction applications. Two high-energy electron ($\gamma>50$) bunches, 5~ns apart, impinge upon a single-foil or a multifoil radiator and generate THz radiation and X-rays simultaneously. The THz pulse from the first bunch is synchronized to the X-ray beam of the second bunch by using an adjustable optical delay of THz pulse. The peak power of THz radiation from the multifoil radiator is estimated to be 0.14~GW for a 200~pC well-optimized electron bunch. GEANT4 simulations show a carbon foil with thickness of 0.5~--~1.0~mm has the highest yield of 10~--~20~keV hard X-rays for a 25~MeV beam, which is approximately 10$^3$ photons/(keV pC-electrons) within a few degrees of the polar angle. A carbon multifoil radiator with 35 foils (25~$\mu$m~thick each) can generate close to $10^3$ hard X-rays/(keV pC-electrons) within a 2$^\circ$ acceptance angle. With 200~pC charge and 100~Hz repetition rate, we can generate $10^7$  X-rays per 1~keV energy bin per second or $10^5$ X-rays per 1~keV energy bin per pulse.	The longitudinal time profile of X-ray pulse ranges from 400~--~600~fs depending on the acceptance angle. The broadening of the time duration of X-ray pulse is observed owing to its diverging effect. A double-crystal monochromator (DCM) will be used to select and transport the desired X-rays to the sample. The heating of the radiators by an electron beam is negligible because of the low beam current.
\end{abstract}
	%
%Valid PACS numbers may be entered using the \verb+\pacs{#1}+ command.
% Classification Scheme.
%Transition radiation by relativistic moving charges, 41.60.Dk
% X-ray beams, 41.50.+h
%linear accelerator, 29.20.Ej

%Use showkeys class option if keyword %display desired
\maketitle
\section{INTRODUCTION}
Ultrafast and high-power terahertz (THz) pulses have been used for applications in chemistry and security inspection~\cite{Tonouchi, Doi}. Such THz radiation at gigawatt peak powers can be achieved by using a multifoil cone radiator~\cite{Vinokurov}. The combination of an ultrafast X-ray beam with such THz pulses, i.e. a THz pump and X-ray probe, is a powerful tool to study the structural response of materials, in a time-resolved manner, to the THz radiation.

We describe a novel source design that generates high power THz pump pulse and X-ray probe pulse by using a 25~MeV electron linac, which is located at KAERI's (Korea Atomic Energy Research Institute's) Radiation Center for Ultrafast Science. In our scheme, two electron bunches, 5~ns apart, generated from a radio-frequency (RF) photogun, are accelerated up to 25~MeV by using an S-band accelerating cavity and are then delivered to a THz/X-ray target. The RMS bunch length at the gun cathode is 1.3~ps and compressed to hundreds of fs (at target) by using a 90$^\circ$ achromatic bend. Electron bunches incident on thin foil(s) generate THz radiation and X-rays, simultaneously. The THz pulse from the first electron bunch is delayed so that it reaches the sample at the same time as the X-ray pulse from the second bunch. By adjusting the arrival time of THz pulse and gating the detectors for the probe signals, we will provide THz-pump and X-ray-probe setup for time-resolved pump-probe experiments.  

\begin{figure}
	\begin{tabular}{cc}
		{\scalebox{0.08} [0.08]{\includegraphics{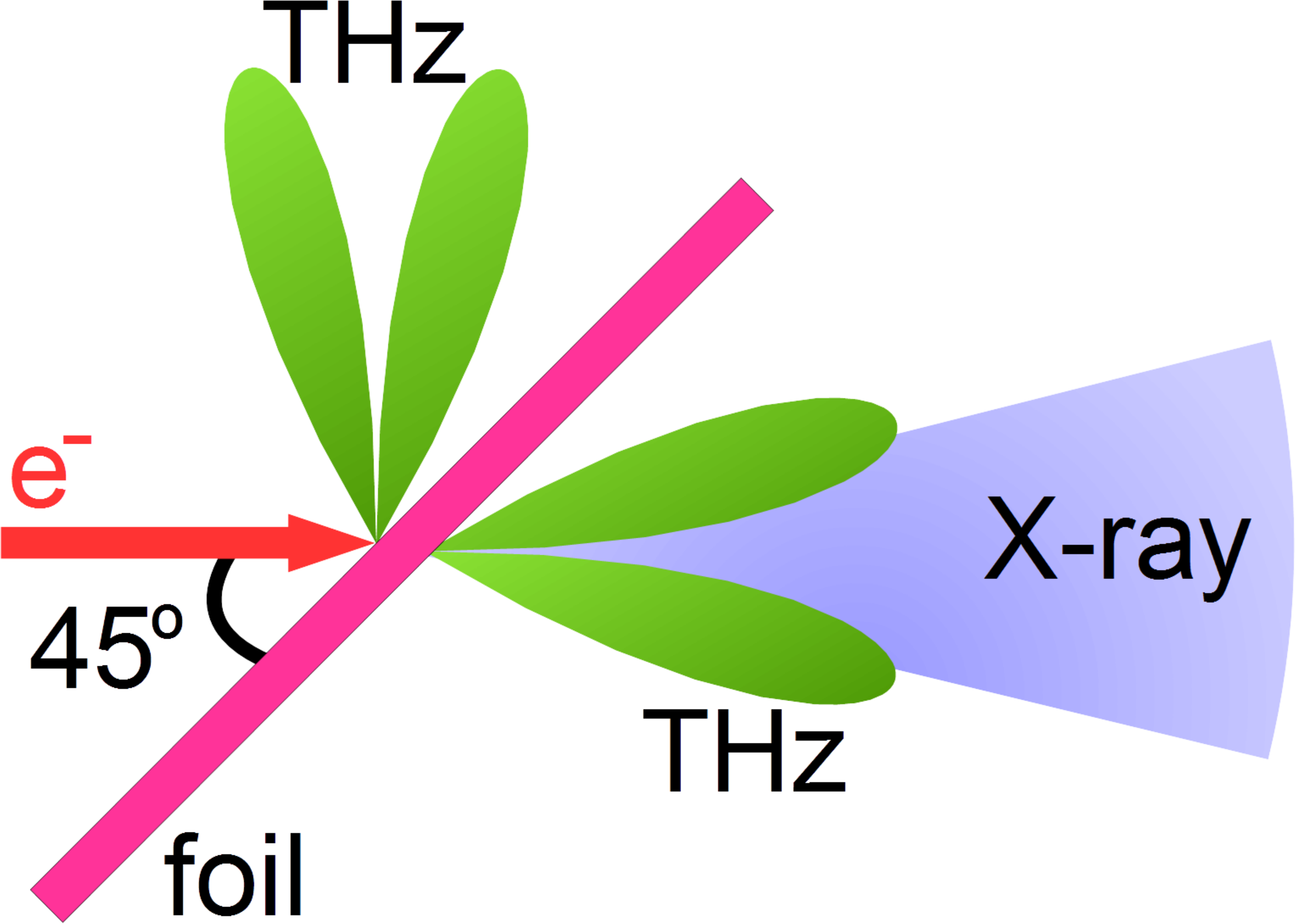}}} ~~&~~ {\scalebox{0.2} [0.2]{\includegraphics{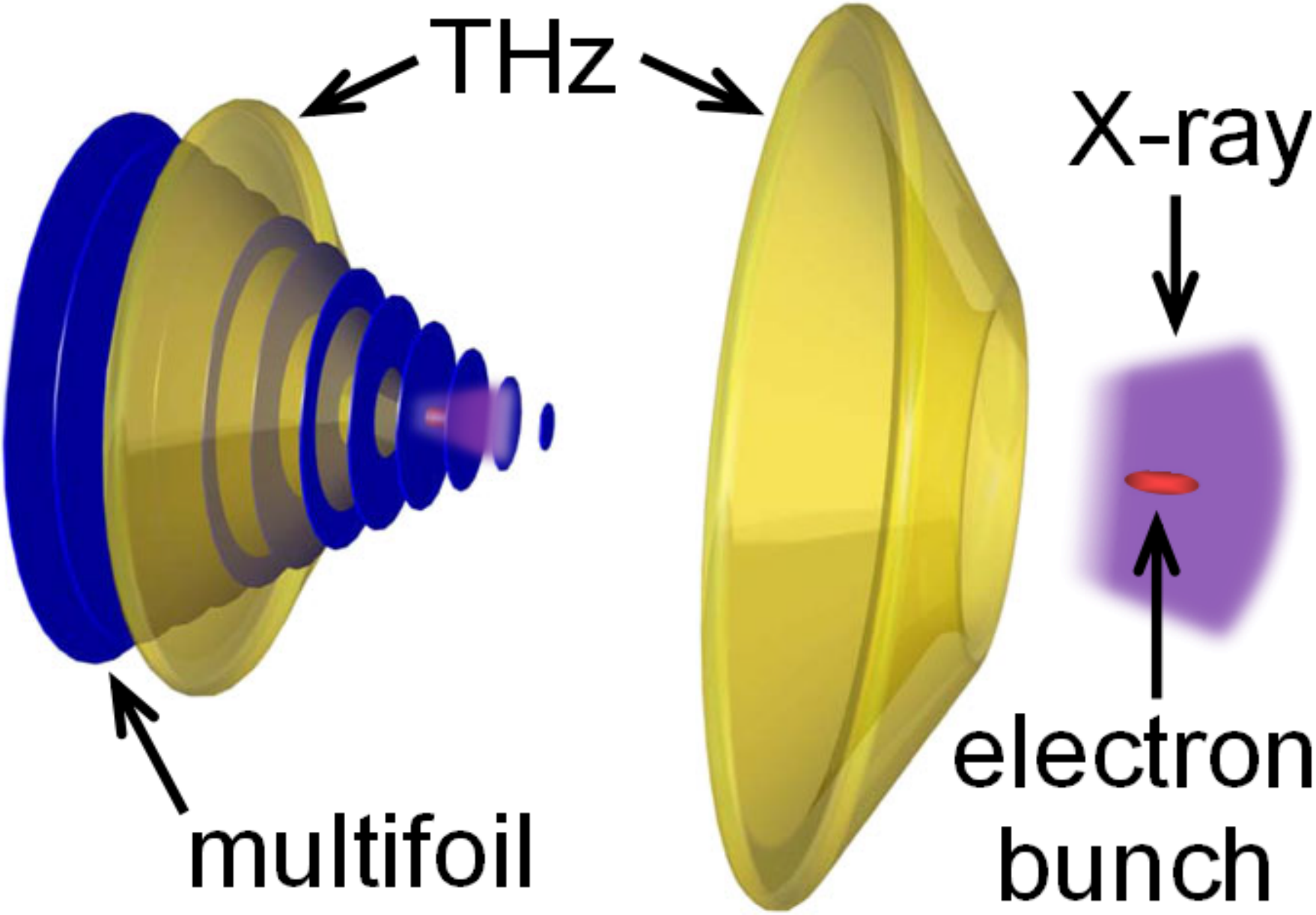}}}\\
		(a) ~~&~~ (b)
	\end{tabular}
	\caption{THz and X-ray generation schemes by using: (a) a single foil radiator, and (b) multifoil radiator~\cite{Vinokurov}. The single foil is placed at a 45$^\circ$ angle with respect to the beam axis.}
	\label{singlefoil}
\end{figure}

We will use two schemes to generate the THz radiation: a single-foil radiator used for testing and diagnostic purposes, and a multifoil for generating high-power THz pulses. In the case of the single foil radiator, as shown in Fig.~\ref{singlefoil}~(a), when ultrashort electron bunches pass through the interface between two media (such as, vacuum and foil), the THz radiation is generated by coherent transition radiation (CTR)~\cite{Daranciang, Hoffmann}. When the foil is placed at a 45$^\circ$ angle with respect to the electron beam propagation, the THz radiation is generated from the front side of the radiator in a cone shape perpendicular to the electron beam axis. The generated THz radiation is collimated by a concave mirror and transported to the sample target being probed. The bremsstrahlung radiation is generated along the direction of the electron beam. 

In the case of a multifoil radiator, as shown in Fig.~\ref{singlefoil}~(b), multiple thin conducting foils with decreasing radii are stacked together to form a truncated cone shaped radiator~\cite{Vinokurov}. Relativistic electron bunches traversing along the cone axis emit THz radiation into each gap between every pair of adjacent foils, and THz radiation propagates radially outward. The radii of the radiators are decreased to synchronize the radiation from each gap to be coherently superposed at the cone edge where an expanding conical THz wave is formed and propagates forward. A THz collimator described in Ref.~\cite{Jo} will be used to collect the THz conical wave and collimate it into a ring-shaped THz beam that travels along the electron beam axis. A mirror with a hole at the center will be used to select the THz beam to the THz delay line while allowing the bremsstrahlung X-rays and residual electron beam to pass through the hole, as shown in Fig.~\ref{wci-layout}. There is a significantly intensive soft X-ray transition radiation (XTR)~\cite{Chu}, but we are not planning to use it. A dipole magnet will be placed beyond the mirror to deflect the residual electrons toward the beam dump.

\section{Schematic of THz Pump/X-ray Probe Beamline}
\begin{figure}
	\includegraphics*[width=85mm]{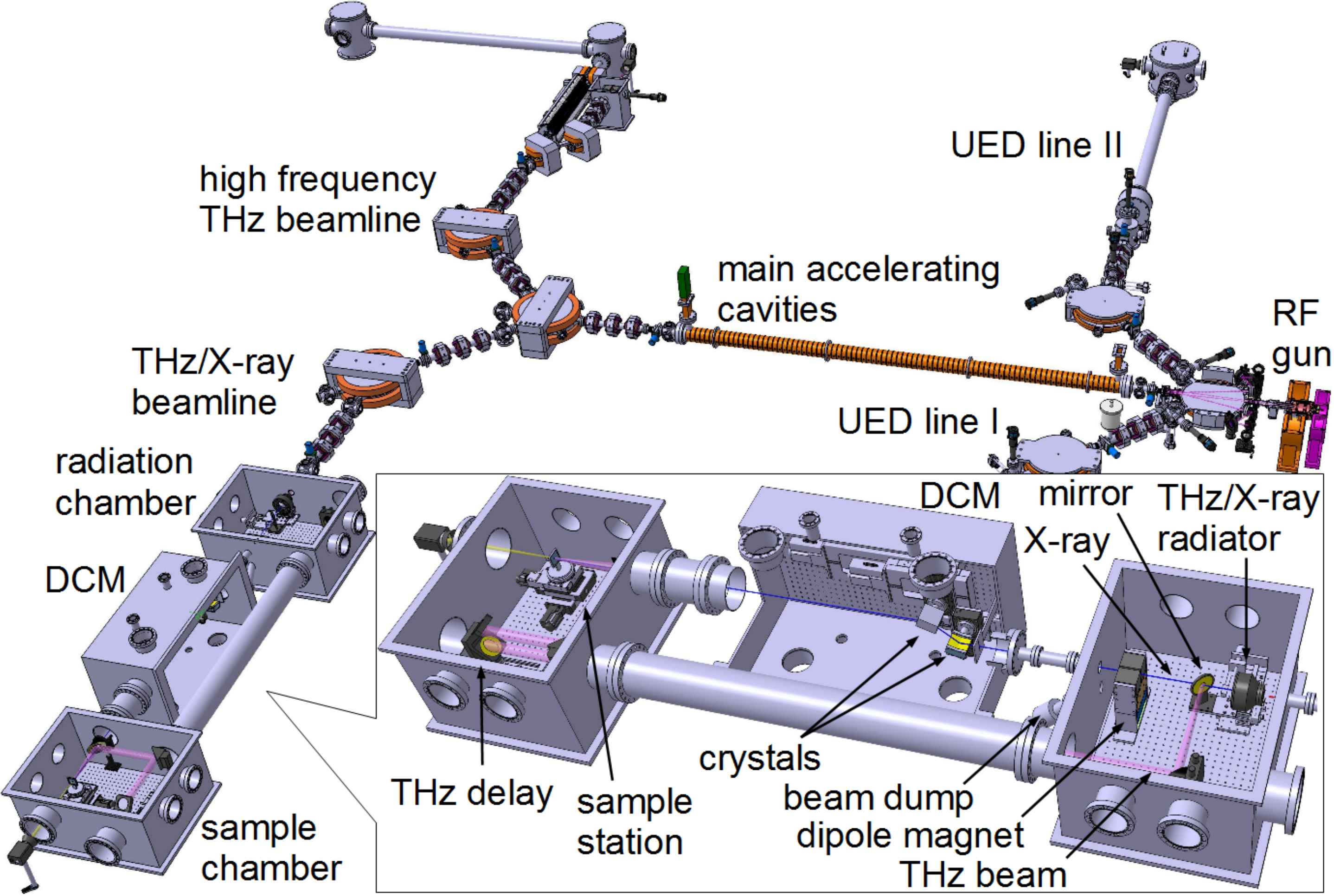}
	\caption{Electron beamline and THz/X-ray pump probe beamline layout of the KAERI Radiation Center for Ultrafast Science.}
	\label{wci-layout}
\end{figure}

The beamline layout is shown in Fig.~\ref{wci-layout}. The electron beam from the gun can be deflected for ultrafast electron diffraction (UED) experiments or further accelerated for THz pump and X-ray probe experiments by a 3-m-long S-band (2856-MHz) linac. The THz/X-ray beamline generates both THz and X-rays by using the target that described above. The high-frequency THz beamline generates THz pulses, by using variable period undulator~\cite{Mun2014}, that will be used as pump or probe sources for investigating materials. The emittance and Twiss parameters of the electron beam out of the gun are measured by using a quadrupole scanning method~\cite{Setiniyaz}, which agrees well with the estimated values~\cite{Kim}. Long-term-stable 10-fs-level synchronization was achieved between the Ti:sapphire photocathode laser and the S-band RF oscillator~\cite{Yang2017}. 

\begin{figure}
	\begin{tabular}{cc}
		{\scalebox{0.167} [0.167]{\includegraphics{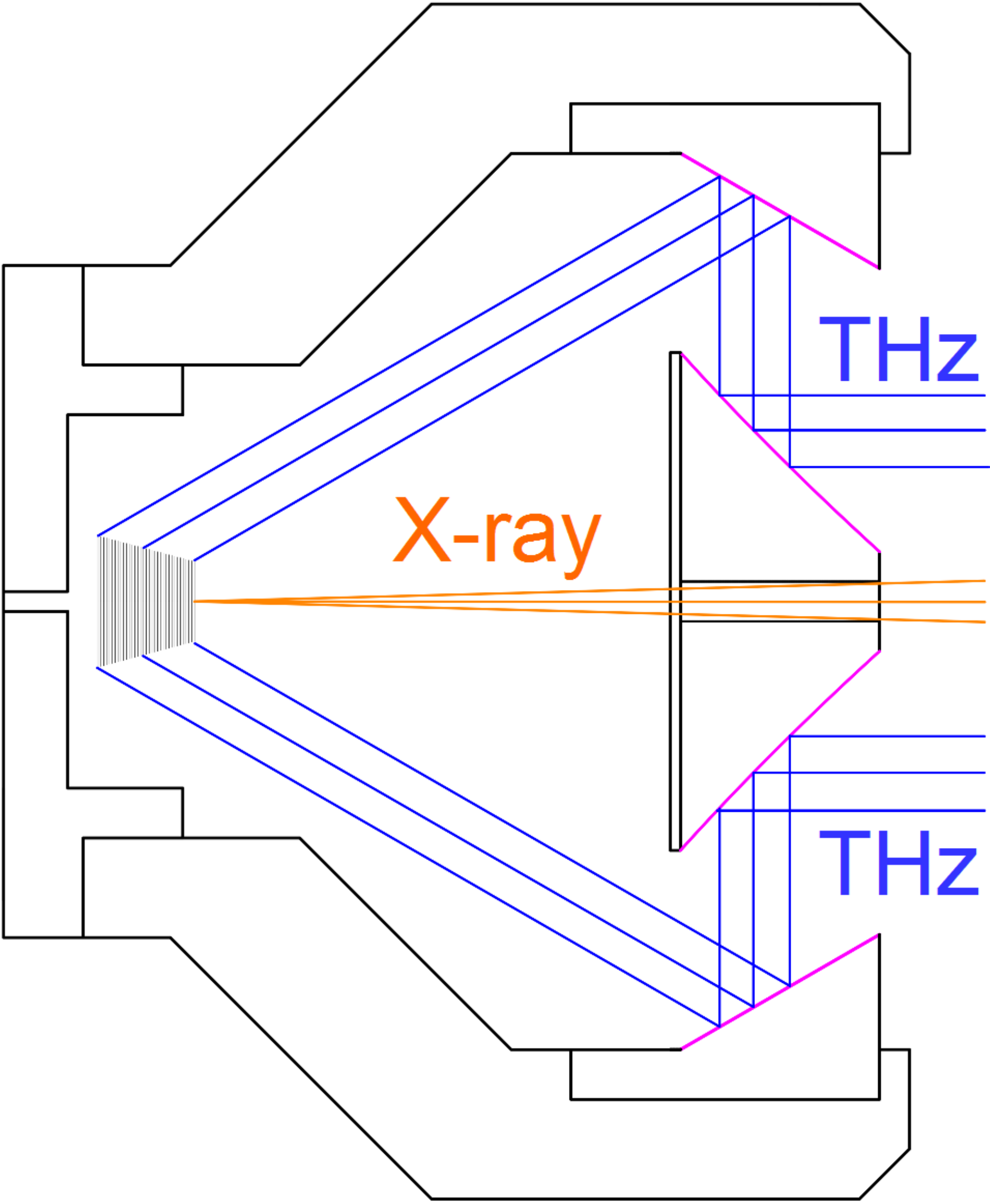}}}~~~&~~~ {\scalebox{0.06} [0.06]{\includegraphics{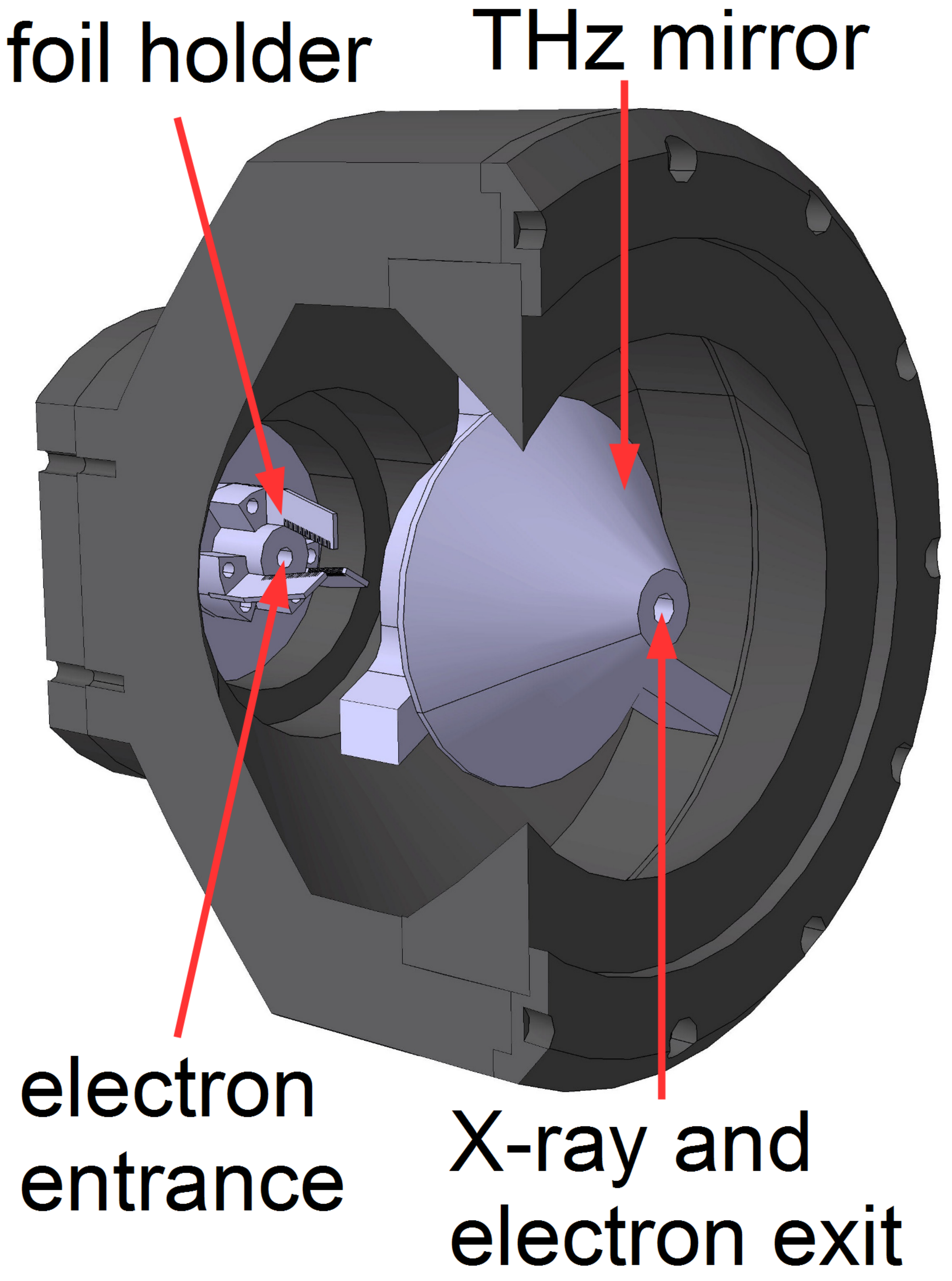}}}\\
		(a) ~~~&~~~ (b)\\	
	\end{tabular}
	\caption{Drawing of the THz and X-ray radiator. (a) Section view. The THz radiations (in blue) are collimated by the THz mirrors (in pink). X-rays (in orange) pass through the exit at the center of the THz mirror. (b) Three dimensional drawing. }
	\label{Radiator}
\end{figure}

The current design of the pump and probe beamline consists of a radiation chamber, a double crystal monochromator (DCM)~\cite{Ishikawa200542, YANG1992422, MORRISON1988467}, and a sample chamber. The radiation chamber hosts THz/X-ray radiators, THz mirrors, a dipole magnet, and a beam dump. An X-ray focusing optics under design will also be installed in the radiation chamber. The DCM utilizes two parallel crystals to select and transmit the desired monochromatic X-ray beam to the sample chamber. The first crystal is fixed on its axis, but is rotatable depending on the X-ray energy. The second crystal can be rotated and translated accordingly. The THz delay and sample station are located in the sample chamber, where the pump and probe experiment is carried out. The drawing of the THz and X-ray radiator is shown in Fig.~\ref{Radiator}. The THz radiation will be reflected by the wall of the radiator holder and the conical mirror to produce a hollow parallel beam. 

\section{THz Radiation}
The THz radiation generated by using a multifoil radiator is described in Ref.~\cite{Vinokurov}. However, unlike the cigar-beam (a long narrow bunch) in the examples, our beam's bunch length is shorter than (or comparable to) its transverse beam size. Therefore, to calculate the THz power for our case, where the foil gaps are large compared to the bunch length, we need to make some modifications to the equations. From Eq.~(3) and Eq.~(6) in Ref.~\cite{Vinokurov}, we can derive
%\begin{widetext}
\begin{equation}
\begin{aligned}
E(r,t) = & -\frac{2(2\pi)^2}{cg}\textnormal{Re}\int\limits_{0}^{\infty}\textnormal{sin}\biggl(\frac{kg}{2}\biggr)\int\limits_{0}^{\infty} j_{\omega}(r) J_{0}(kr)rdr \\
& \times H_{0}^{(1)}(kr) e^{-i \omega t} \frac{d\omega}{2\pi}\\
= & E_{1}\biggl( t + \frac{g}{2c} \biggr) - E_{1}\biggl( t - \frac{g}{2c} \biggr),
\end{aligned}
\label{eq:wideeq2}
\end{equation}
where $c$ is speed of light, $g$ is the foil gap distance, $k=\omega/c$, with $\omega$ being the frequency of the radiation ($n=1$ for vacuum), $j_{\omega}(r)$ is beam current density, $J_{0}(kr)$ and $H_{0}^{(1)}(kr)$ are zeroth-order Bessel and Hankel functions, respectively, and 
\begin{equation}
\begin{aligned}
E_{1}(r,t) = &\frac{(2\pi)^2}{cg}\textnormal{Im}\int\limits_{0}^{\infty} \int\limits_{0}^{\infty} j_{\omega}(r) J_{0}(kr)rdrH_{0}^{(1)}(kr) \\
& \times e^{-i \omega t} \frac{d\omega}{2\pi}.
\end{aligned}
\label{eq-E2}
\end{equation}
This formula shows that the radiation field in the gap consists of two transition radiation pulses that were generated when the bunch enters and leaves the gap. Assuming a Gaussian beam with a charge density of
\begin{equation}
j(t)=I(t)\frac{1}{2\pi a^2}e^{-\frac{r^2}{2a^2}},
\label{eq-j}
\end{equation}
where $a$ is the RMS beam size, one may simplify Eq.~\ref{eq-E2} and get 
\begin{equation}
\begin{aligned}
E_{1}(r,t) = &\frac{2\pi}{cg}\textnormal{Im}\int\limits_{0}^{\infty}  I_{\omega} e^{-\frac{k^2a^2}{2}} H_{0}^{(1)}(kr) e^{-i \omega t}  \frac{d\omega}{2\pi} \\
\approx &~-\frac{\sqrt{2}}{g\sqrt{cr}} \int\limits_{0}^{\infty} I_{1}(t-\tau-r/c) \frac{d\tau}{\sqrt{\tau}},
\end{aligned}
\label{eq-E4}
\end{equation}
where 
\begin{equation}
\begin{aligned}
I_{1}(t) = & \int\limits_{-\infty}^{\infty} I_{\omega} e^{-\frac{k^2a^2}{2}} e^{-i \omega t} \frac{d\omega}{2\pi} \\
= & \int\limits_{-\infty}^{\infty}  I(s) e^{-\frac{1}{2a^2}c^2(t-s)^2}\frac{c ds}{\sqrt{2\pi}a}. \\
\end{aligned}
\label{eq-It0}
\end{equation}
Let
\begin{equation}
I(t)=\frac{Q}{N}\sum_{n=1}^{N}\delta(t-T_{n}),
\label{eq-It}
\end{equation}
where $T_{n}$ and $Q$ is the particle's time and charge, respectively. Then 
\begin{equation}
\begin{aligned}
I_{1}(t) %= & \int\limits_{-\infty}^{\infty}  I(s) e^{-\frac{1}{2a^2}c^2(t-s)^2}\frac{c ds}{\sqrt{2\pi}a} \\
= & \frac{Q}{N}\sum_{n=1}^{N} \frac{c}{\sqrt{2\pi}a} e^{-\frac{1}{2a^2}c^2(t-T_{n})^2}
\end{aligned}
\label{eq-I}
\end{equation}
and
\begin{equation}
\begin{aligned}
E_{1}(r,t)
%\approx &~-\frac{\sqrt{2}}{g\sqrt{cr}} \int\limits_{0}^{\infty} I_{1}(t-\tau-r/c) \frac{d\tau}{\sqrt{\tau}}\\
%= & -\frac{\sqrt{2}Q}{g\sqrt{ar}N} \sum_{n=1}^{N} \int\limits_{0}^{\infty} \frac{1}{\sqrt{2\pi}}e^{-\frac{1}{2} \bigl(c\frac{t-r/c-T_{n}}{a}-y\bigr)^2 } \frac{dy}{\sqrt{y}}\\
\approx &~ -\frac{\sqrt{2}Q}{g\sqrt{ar}N} \sum_{n=1}^{N} F\biggl( c\frac{t-r/c-T_{n}}{a} \biggr),
\end{aligned}
\label{eq-E5}
\end{equation}
where 
\begin{equation}
F\biggl( c\frac{t-r/c-T_{n}}{a} \biggr)=\int\limits_{0}^{\infty} \frac{1}{\sqrt{2\pi}}e^{-\frac{1}{2} \bigl(c\frac{t-r/c-T_{n}}{a}-y\bigr)^2 } \frac{dy}{\sqrt{y}}
\label{eq-E6}
\end{equation}
and $y \equiv \frac{c\tau}{a}$. Hence, for a short (with respect to $a$) electron bunch, the field is
\begin{equation}
\begin{aligned}
E(r,t) = & -\frac{\sqrt{2}Q}{g\sqrt{ar}N} \Biggl[ F\biggl( \frac{ct-r+g/2}{a} \biggr) \\
& - F\biggl( \frac{ct-r-g/2}{a} \biggr) \Biggr].
\end{aligned}
\label{eq-E7}
\end{equation}
For a simple estimation, when the beamsize is small compared to the gap ($g>>a$), the maximum field is approximately 
\begin{equation}
\begin{aligned}
|E_{1}|_{\textnormal{max}} %= & \Bigg[ \frac{\sqrt{2}}{g\sqrt{cr}} \int\limits_{0}^{\infty}  I_{1}(t-\tau-r/c) \frac{d\tau}{\sqrt{\tau}} \Bigg]_{\textnormal{max}} \\
\approx &~\frac{\sqrt{2}Q}{g\sqrt{r\sigma}},
\end{aligned}
\label{eq-Emax}
\end{equation}
where $\sigma=\sqrt{l^2+a^2}$ with $l$ being the RMS bunch length. Then, one may obtain THz peak power by using
\begin{equation}
\begin{aligned}
P_{\textnormal{max}}= &\frac{c}{4\pi} |E_{1}|_{\textnormal{max}}^{2} 2 \pi r L = \frac{cQ^{2}}{g^{2}\sigma}L,  \\
%= & \frac{cQ^{2}}{g^{2}\sqrt{l^2+a^2}}L.
\end{aligned}
\label{eq-THz-Power}
\end{equation}
where $L$ is the multifoil cone height. Eq.~\ref{eq-THz-Power} shows that when bunch charge and multifoil geometry are set, the radiated THz power peak power is equally determined by bunch length and beamsize.

\section{Beam Optimization and Transportation Simulations}
Currently, the main accelerating cavity is under conditioning and the electron beam parameters at the THz radiators are obtained by using ASTRA~\cite{ASTRA} and ELEGANT~\cite{ELEGANT} simulations alternatively in three steps. In the simulations, 2$\times$10$^5$ macro-particles are transported from the gun to radiators. Each macro particle carries 1~fC charge and the total bunch charge is 200~pC.

The first step is the generation and acceleration of the beam from the RF gun to the end of the main accelerating cavities by using ASTRA. The beam energy inside the RF gun is low and ASTRA can simulate the space charge effect which is dominant at low energies. The beam energy out of the gun is 3~MeV and after the main accelerating cavities reaches 25~MeV.

In the second step, from the end of the main accelerating cavities to the end of the second bending dipole, ELEGANT was used. ELEGANT can simulate the CSR (Coherent Synchrotron Radiation) and the ISR (Incoherent Synchrotron Radiation) effects in bending magnets. The result shows that effects of CSR and ISR on the beam are negligible. At this step, the RMS beamsize and RMS bunch length is about 1~mm and 1~ps, respectively. The beam energy is about 25~MeV, therefore, the space charge effect is not significant. Another reason for using ELEGANT is the proper quadrupole strength to suppress dispersion can be found easily by using the ELEGANT parameter scan. 

\begin{figure}
	\begin{tabular}{c}
		\begin{overpic}[width=7cm]{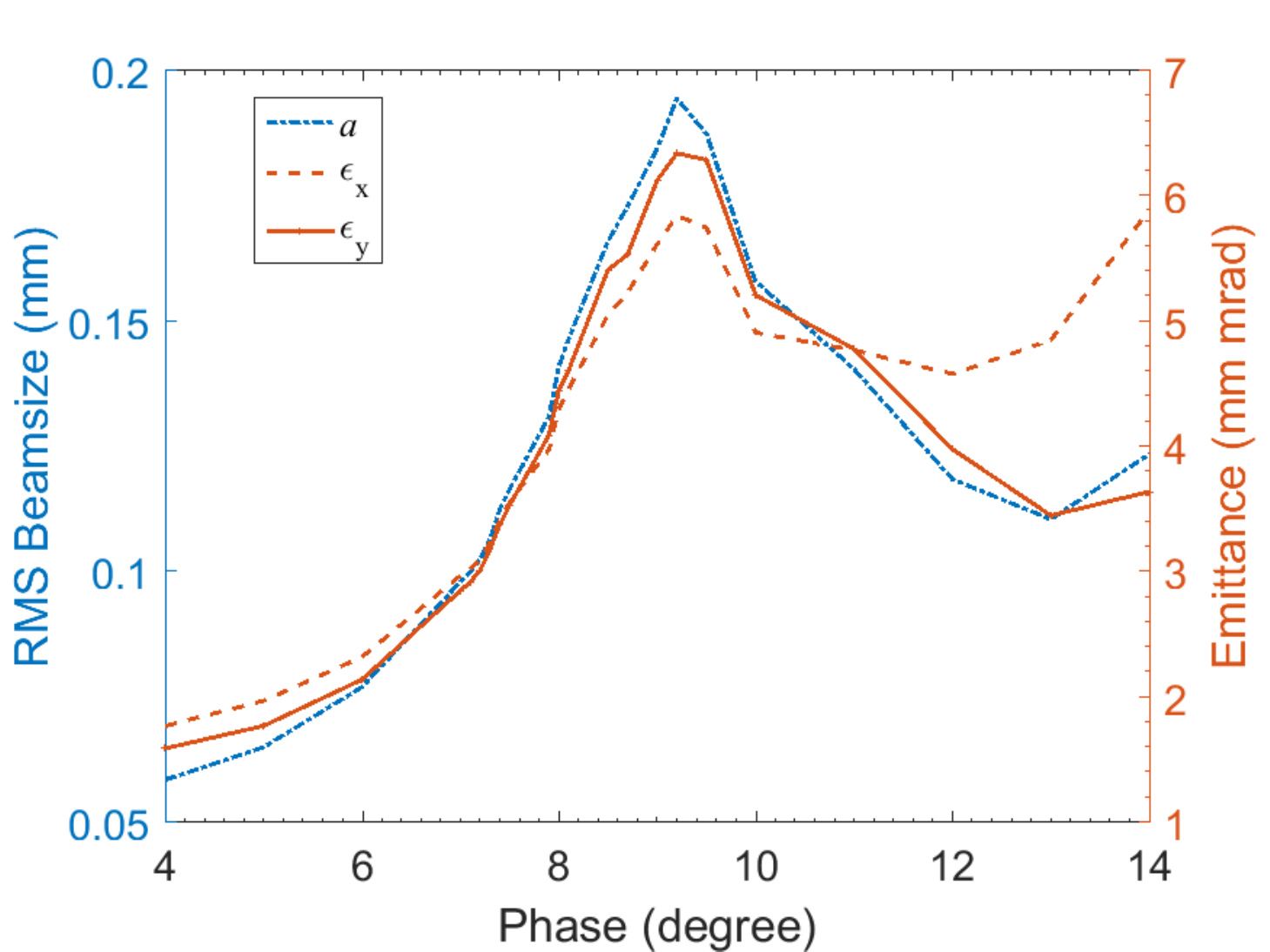} \put(70,60){(a)} \end{overpic}\\
		\begin{overpic}[width=7cm]{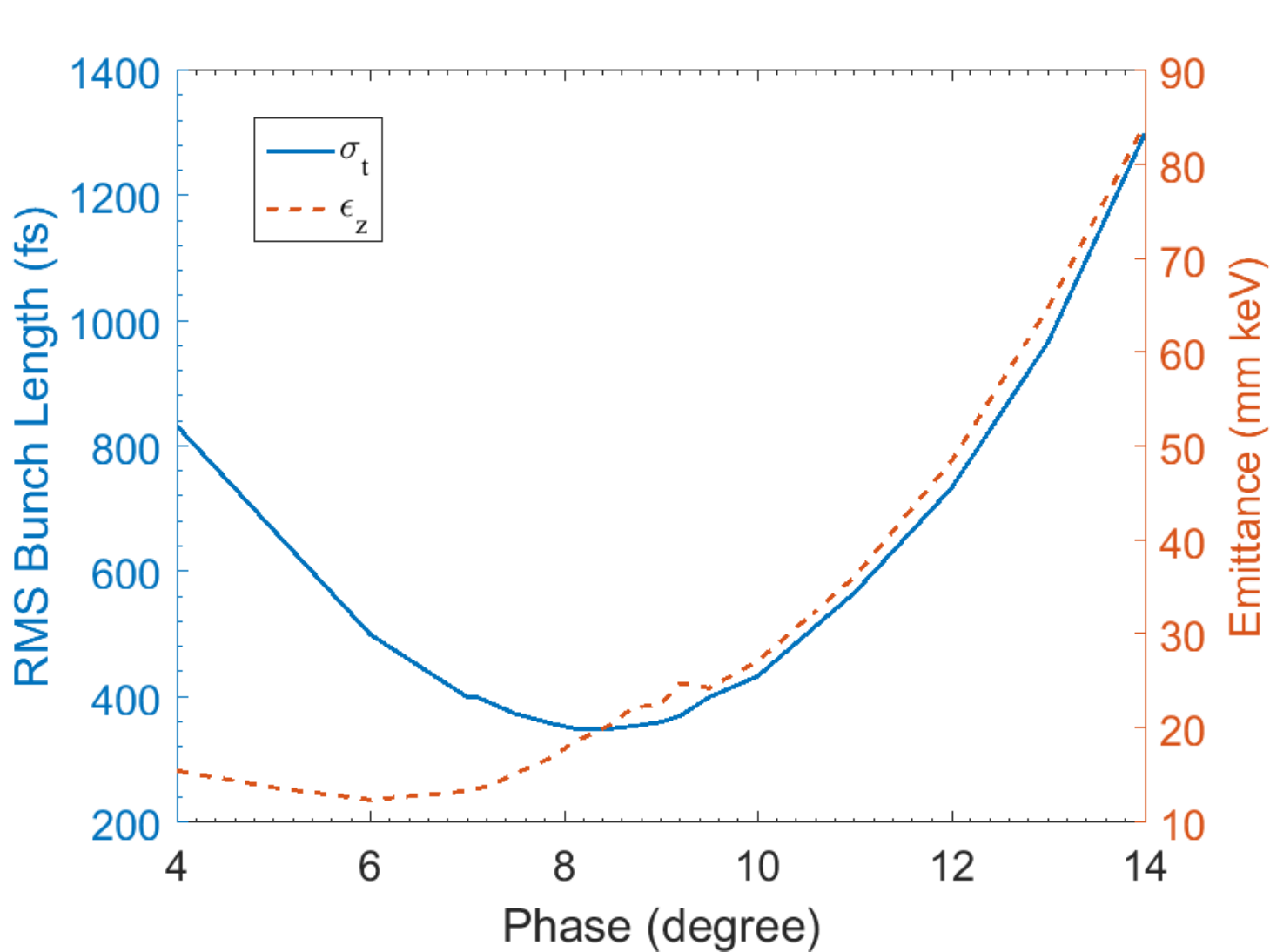} \put(70,60){(b)} \end{overpic}\\
	\end{tabular}
	\caption{RF phase scan results. (a) Transverse RMS beamsize and emittances vs. RF phase. (b) RMS bunch length and longitudinal emittance vs. RF phase.}
	\label{PhaseScan}
\end{figure}

In the last step, from the end of the second bending dipole to the end of the multifoil radiator, ASTRA was used. This is because as the beam is focused and compressed, therefore the space charge effect starts to dominate. This can be seen from the beam RF phase scan results shown in Fig.~\ref{PhaseScan}. When the RF phase increases, the bunch lengths shortens and the space charge force becomes dominant, which causes the emittances (both longitudinal and transverse) and transverse beamsize to increase. The maximum compression of the bunch length happens around 9$^\circ$ RF phase, and the corresponding transverse emittances and beamsizes are also largest. When RF phase is larger than 9$^\circ$, the longitudinal compression weakens, so the growth of transverse beamsizes and emittances are reduced. 

\begin{figure}
	\centering
	\includegraphics[width=0.9\linewidth]{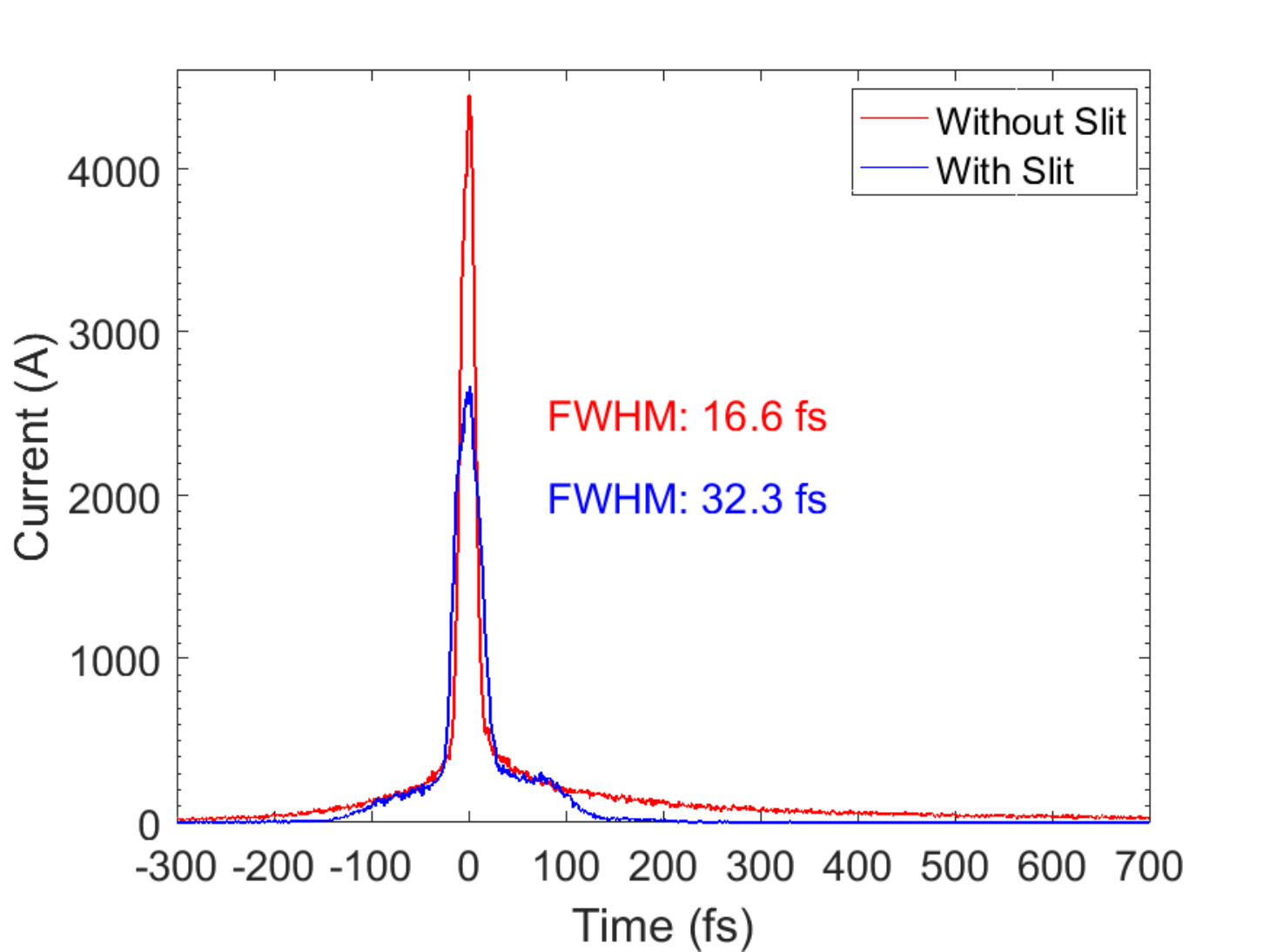}
	\caption{Electron beam current at the target with (blue line) and without (red line) the slit in the beamline. The RF phase is 9$^\circ$.}
	\label{fig:WithAndNoSlit}
\end{figure}

The shortest FWHM bunch length of 16.6~fs occurred when RF phase set around 9$^\circ$ as shown by the red curve in Fig.~\ref{fig:WithAndNoSlit}. The corresponding peak current is 4.4~kA. This peak has long tails that are the main contributors of the RMS bunch length and increase it up to 350~fs. They can be cutoff by placing an energy slit in the dispersive section. A 5~mm wide slit was placed in the 6-way cross chamber (placed after the first dipole magnet), but it also decreased beam peak current and increased bunch length, as shown by the blue curve. This could be explained by the missing of the space charge force that pushes the bunch from the two ends and keeps it longitudinally compact. Adding slit to the current setup seems to be counterproductive and further investigation is necessary. 

\begin{figure}
	\centering
	\includegraphics[width=0.9\linewidth]{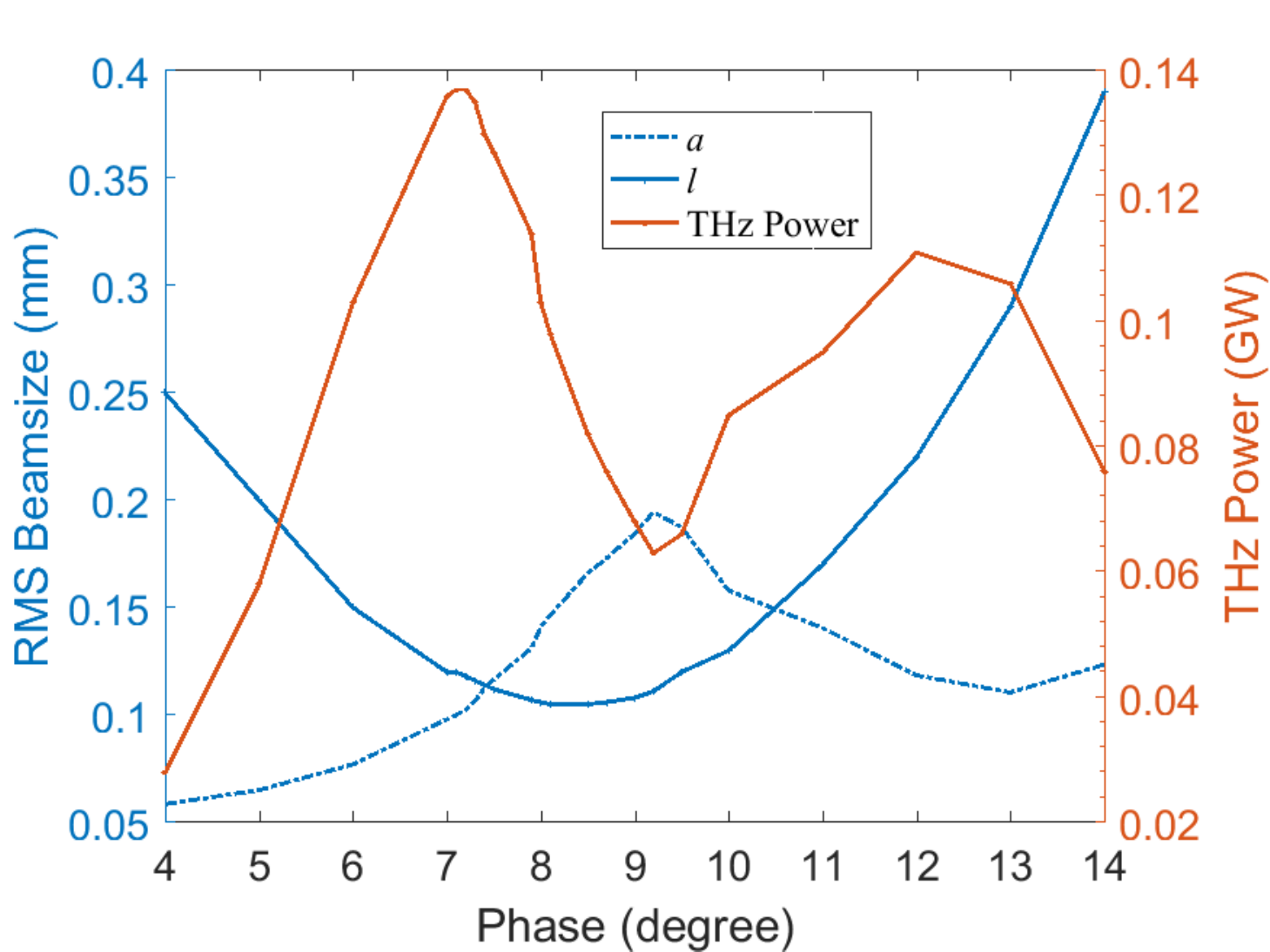}
	\caption{THz peak powers estimated by using Eq.~\ref{eq-E7} for different RF phases.}
	\label{fig:THz-Powers}
\end{figure}

The THz peak powers estimated by using Eq.~\ref{eq-E7} for different RF phases are given in Fig.~\ref{fig:THz-Powers}. The highest peak power is about 0.14~GW, when RF phases is 7.1$^\circ$. Even though the bunch length is shortest at 9$^\circ$, but the beamsize is largest as well and becomes the dominant factor to decrease the THz peak power. So, to maximize THz peak power, the RF phase should be set few degrees off from the 9$^\circ$ RF phase and avoid overgrowth of the beamsize. At 7.1 degree RF phase, the longitudinal and transverse beamsize is the same, therefore the beam volume is most compact and $\sigma=\sqrt{l^2+a^2}$ is at minimum. This result in the highest THz peak power that can be seen in Fig.~\ref{fig:THz-Powers} and can be explained by Eq.~\ref{eq-THz-Power}.

\begin{figure}
	\centering
	\includegraphics[width=0.9\linewidth]{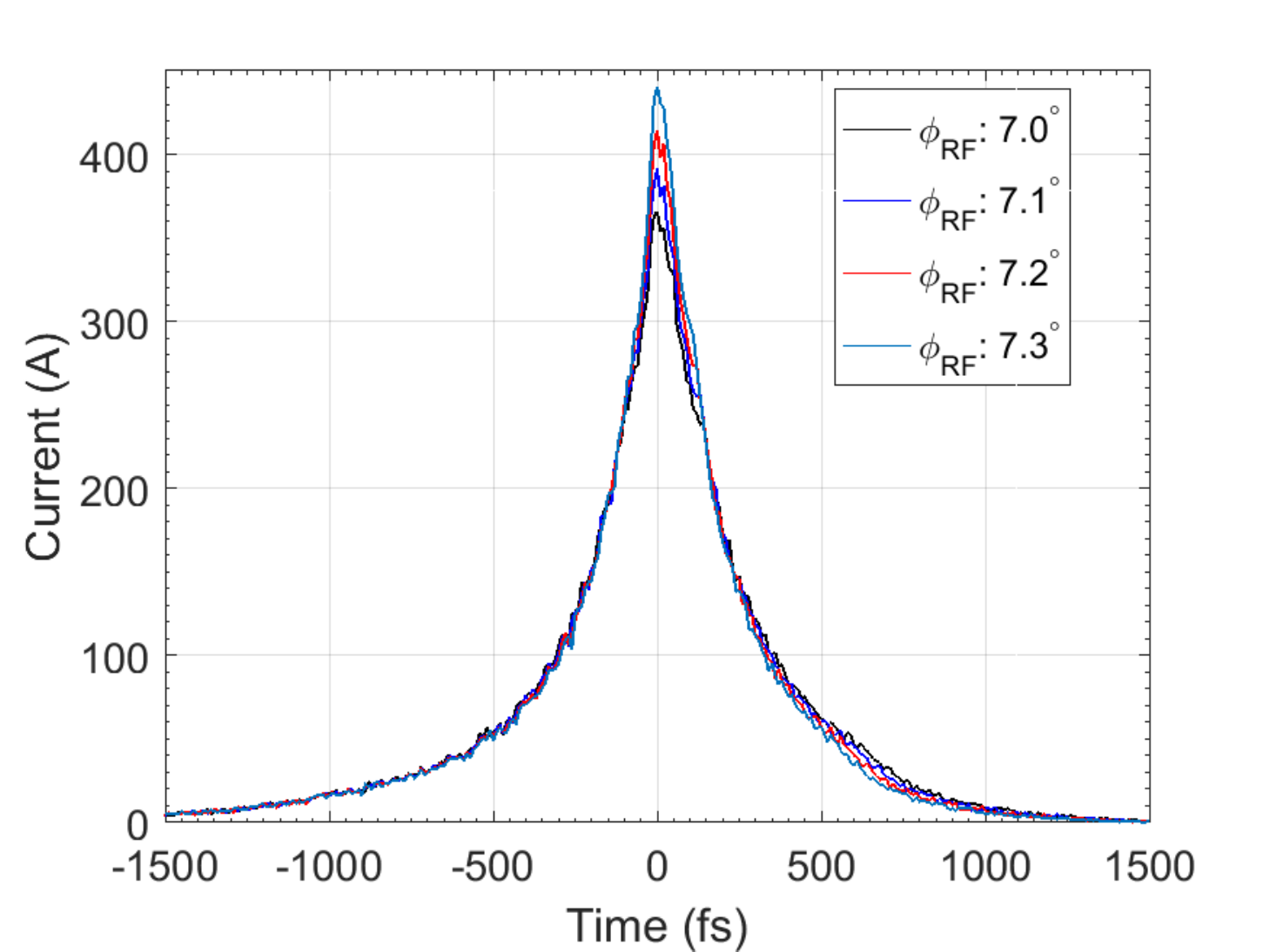}
	\caption{Time profile of the electron beam around 7.1$^\circ$ RF phase.}
	\label{fig:RF_Jitter}
\end{figure}

We performed RF phase scan around 7.1$^\circ$ to estimate the sensitivity of the beam to the RF phase jitter. The results are given in Fig.~\ref{fig:RF_Jitter}. When $\phi_{\textnormal{RF}}$ = 7.1$^\circ$, the peak current is 400~A with RMS bunch length of 330~fs. The measured RMS RF phase and amplitude jitter at the gun are 0.017$^\circ$ and 0.05$\%$ respectively. If we assume similar jitter at main linac, within 3$\sigma_{\textnormal{RF}}$=$0.1^\circ$ range (i.e. with 99.7$\%$ probability), the fluctuation of this peak current is less than 5$\%$. The amplitude jitter creates corresponding energy jitter in the beam, but the its effect on the energy spread is negligible. The compression of the bunch length is mainly determined by the energy spread, therefore amplitude jitter has no significant effect the peak current. The timing jitter is still under research and beyond scope of this paper. 

\section{X-Ray Radiator Optimization}
\subsection{X-rays from Single Foil}
X-rays are generated via the bremsstrahlung process when the high energy electrons traverse the radiator. The intensity of the X-rays is proportional to the square of the atomic number, Z$^2$, of the radiator material and the thickness of the radiator foil~\cite{RevModPhys.31.920}. However, these parameters are closely related to the attenuation of the X-rays~\cite{JACKSON1981169, HUBBELL19821269}. One may optimize the radiator material and thickness for a specific X-ray energy. 

\begin{figure}
	\includegraphics*[width=28mm]{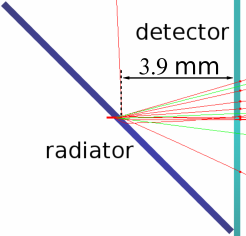}
	\caption{GEANT4 simulation setup for radiator optimization.}
	\label{sim-setup}
\end{figure}

\begin{figure*}
	\begin{tabular}{cc}
		{\scalebox{0.55} [0.55]{\includegraphics{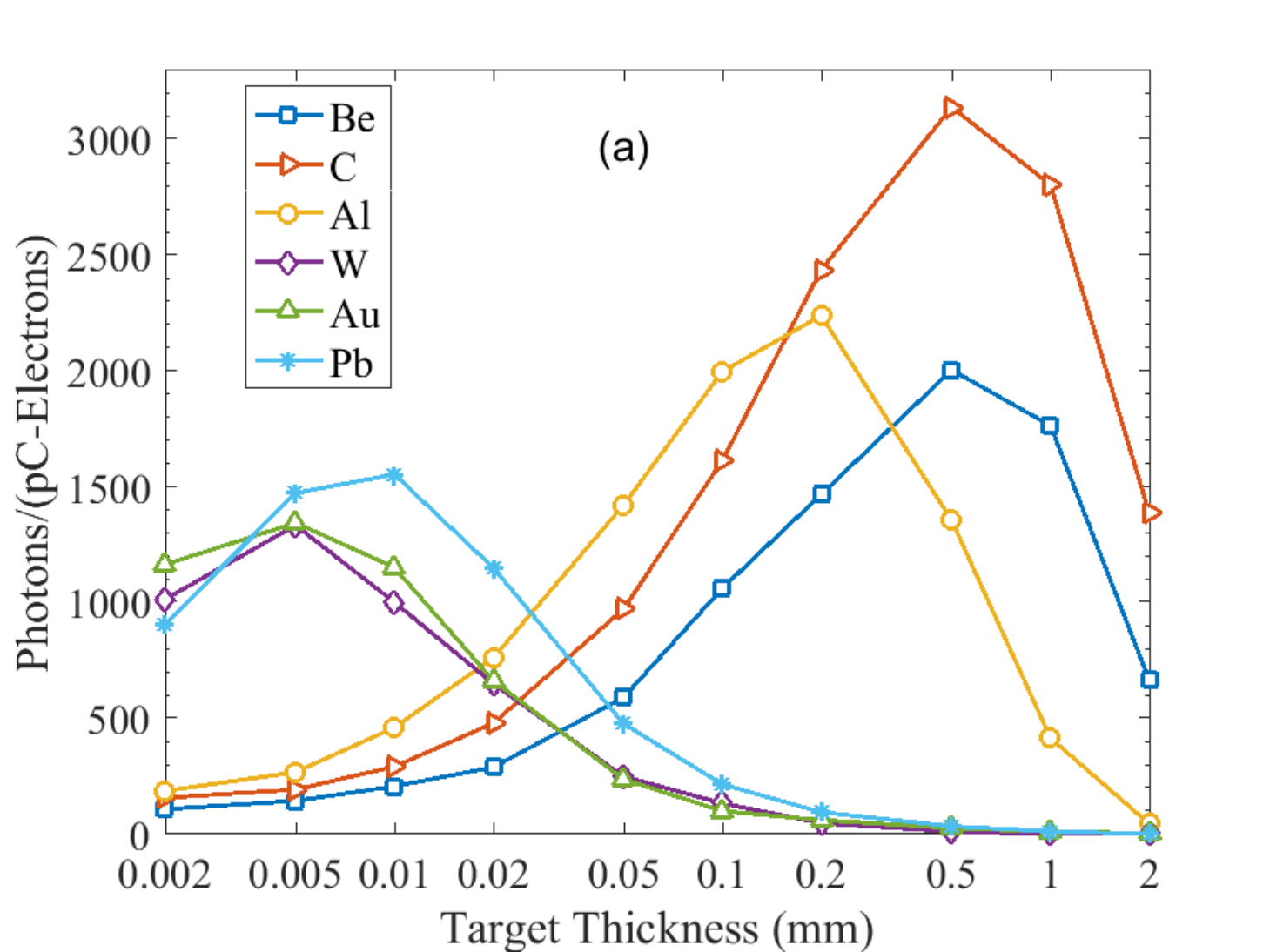}}} ~~~&~~~ {\scalebox{0.55} [0.55]{\includegraphics{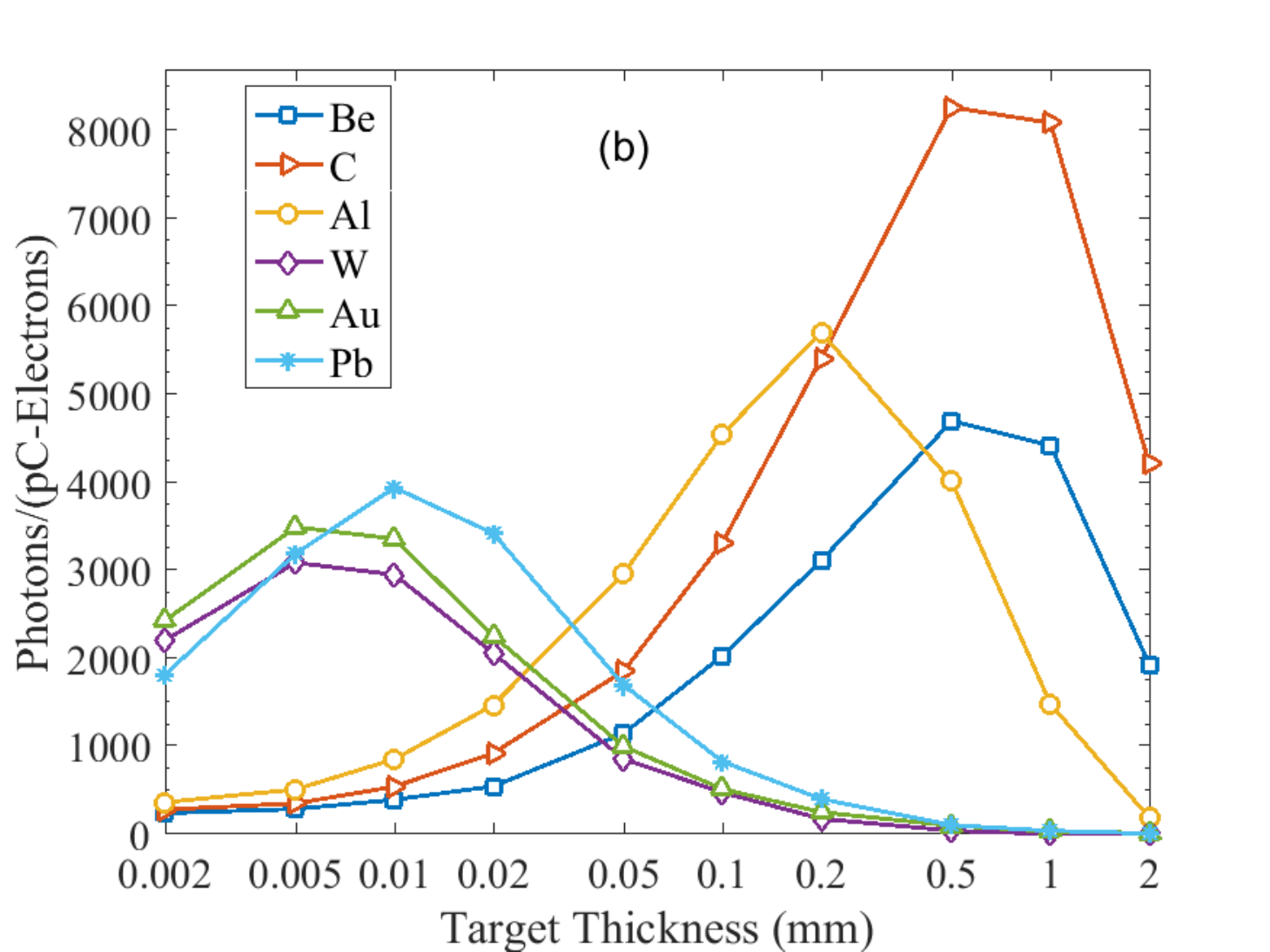}}}\\	
		{\scalebox{0.55} [0.55]{\includegraphics{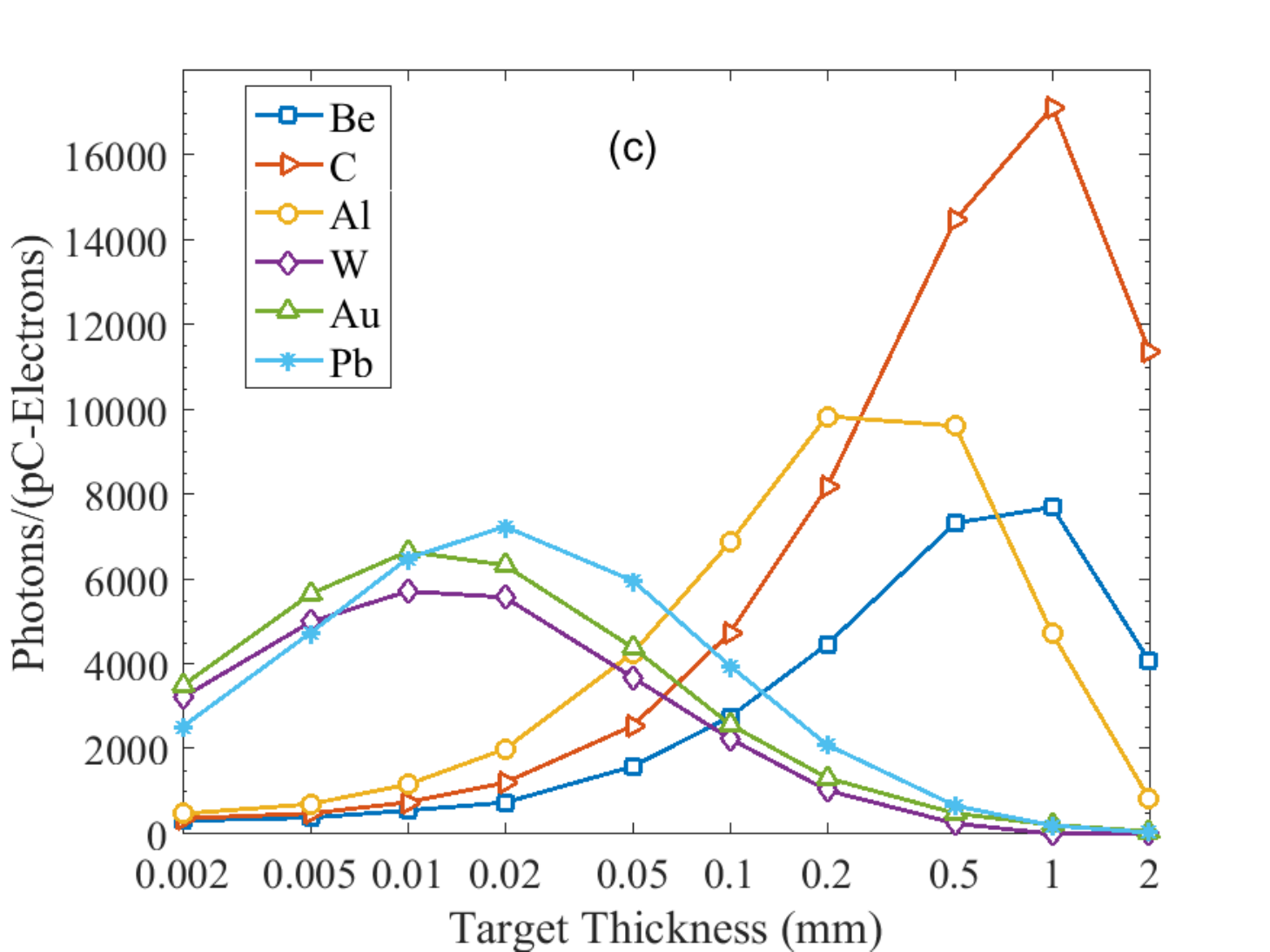}}} ~~~&~~~ {\scalebox{0.55} [0.55]{\includegraphics{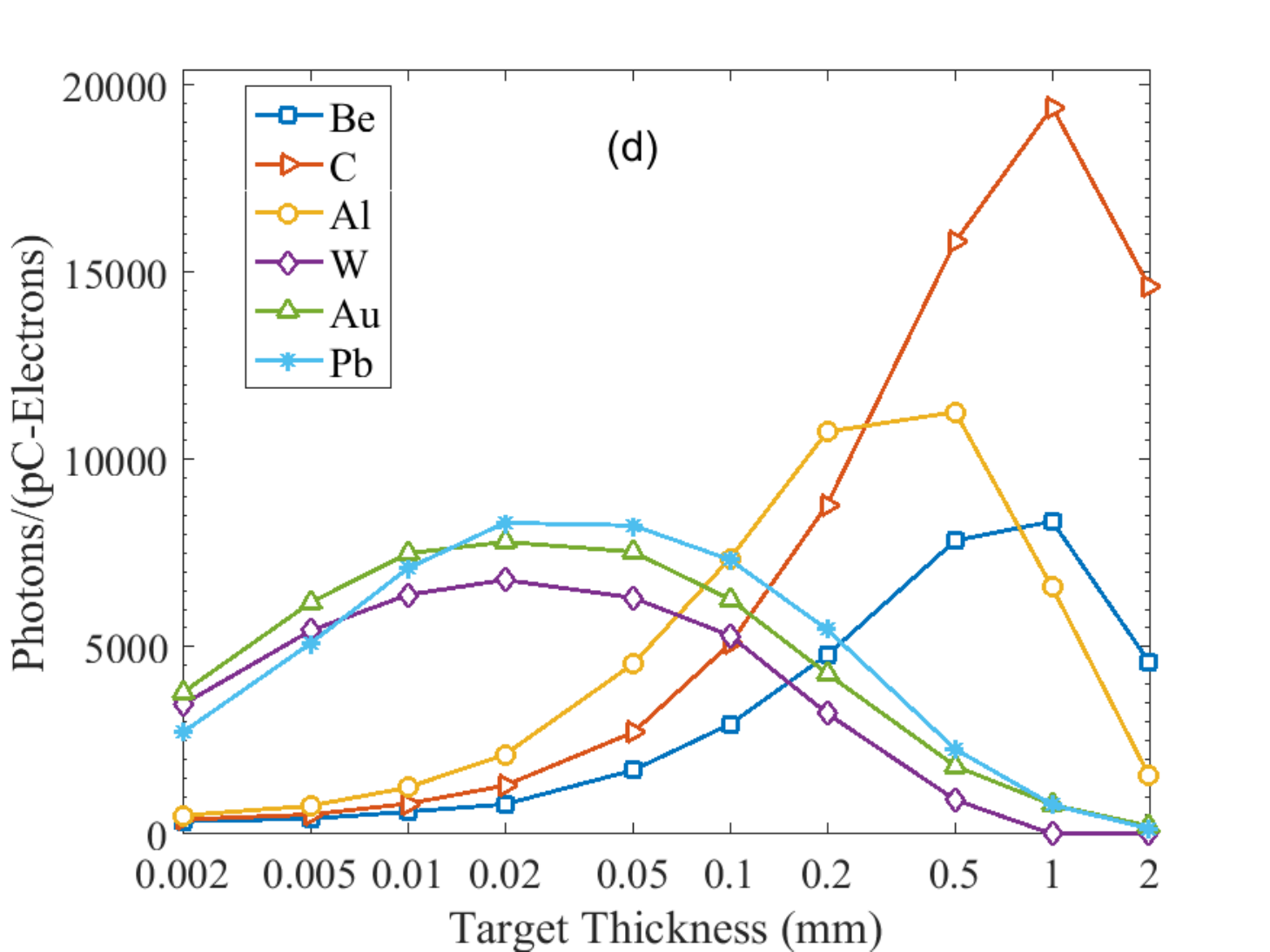}}}\\
	\end{tabular}
	\caption{The number of 10~--~20~keV photons generated by the 1~pC electron beam incident on foils with different thicknesses and materials. The electron beam parameters are given in Table~\ref{sim-para}. The subplots show photons with polar angles ($\theta$) smaller than: (a) 1$^\circ$, (b) 2$^\circ$, (c) 5$^\circ$, and (d) 10$^\circ$.}
	\label{X-ray-counts}
\end{figure*}

Simulations were performed to search for the radiator material with the highest yield of X-rays in the energy range of 10~--~20~keV by using GEANT4~\cite{GEANT4}. The low energy processes in GEANT4 were activated. Note that the XTR (X-ray Transition Radiation) process is not included in the simulation because it generates very low energy X-rays, which are not of interest to us. The incident beam parameters, obtained by using ASTRA/ELEGANT simulations, used in the GEANT4 simulation were shown in Table~\ref{sim-para}. In this study, we group X-rays according to the polar angles, so the size of beam is irrelevant. In the simulation, 1~pC (i.e. 6.25 million) electrons were fired at the radiator. The radiator used in the simulation are beryllium, carbon, aluminum, tungsten, gold, and lead with thicknesses of 0.002, 0.005, 0.01, 0.05, 0.1, 0.2, 0.5, 1.0, and 2.0~mm. As shown in the simulation setup in Fig.~\ref{sim-setup}, the radiator target (blue) is placed at a 45$^\circ$ angle with respect to the beam axis, which is similar to the single foil radiator scheme shown in Fig.~\ref{singlefoil}~(a). A detector (cyan) is placed 3.9~mm beyond the target center perpendicular to the beam axis. The diameter of the radiator and detector are 10.9 and 20~mm, respectively. The red/green lines are electron/photon trajectories. The number of 10~--~20~keV photons with polar angles smaller than 1$^\circ$, 2$^\circ$, 5$^\circ$, and 10$^\circ$ are shown in Fig.~\ref{X-ray-counts}. 
\begin{table}
	\caption{The incident electron beam parameters used in GEANT4 simulation for single foil radiator.}
	\begin{ruledtabular}
		\begin{tabular}{lcc}
			{Parameter} & {Unit}  & {Value}    \\		
			 \hline
			beam energy	&  MeV     &  25.0  \\
			RMS energy spread	&    MeV   &  0.3  \\		
			beam size $\sigma_{x}/\sigma_{y}$	&   mm  &  0.097/0.103  \\
			beam divergence $\sigma_{x'}/\sigma_{y'}$	&  mrad  & 0.71/0.69\\
			RMS bunch length	&   mm    &  0.12  \\	
		\end{tabular}
	\end{ruledtabular}
	\label{sim-para}
\end{table}

The overall results show that the carbon foil gives the highest X-ray yield in the energy range of 10~--~20~keV. High Z materials, such as lead, have their peak intensity for a thin foil less than 20~$\mu$m in thickness, while low Z materials, such as carbon, have their peak intensity for a foil thicker than 500~$\mu$m. The maximum yield using carbon foil is almost twice that of obtainable while using lead foils. As the acceptance angle increases, the peak is shifted toward thicker ones, which is due to the scattering effect~\cite{Birkholz2006}. Note that the effective thickness of the beam path is $\sqrt{2}$ times the target thickness.

\begin{figure}
	\includegraphics*[width=70mm]{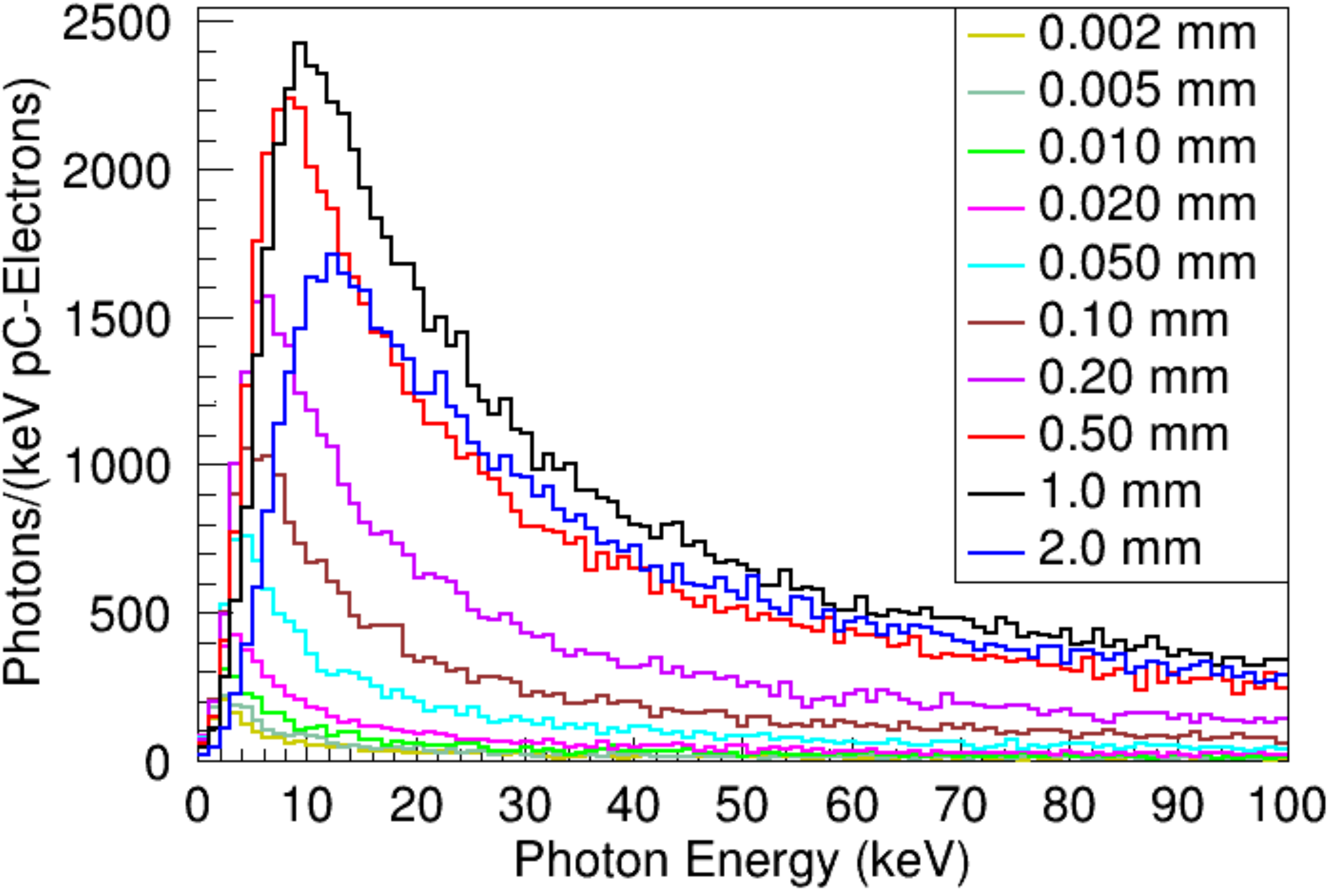}
	\caption{The X-ray energy spectrum observed on the detector generated by 1~pC electrons incident on the various thicknesses of carbon foil. The counts are given in 1~keV energy bins.}
	\label{C-foils-En}
\end{figure}

We investigated the X-rays generated from the carbon single foil radiators. The X-ray energy spectrum generated from the various thicknesses of carbon foil are shown in Fig.~\ref{C-foils-En}. As can be seen, as the foil thickness was increased up to 1.0~mm, the X-ray yield also increased and the peak shifted from left (lower energies around 3~--~4~keV) to right (higher energies around 10~keV). When the foil thickness reached 2.0~mm, the peak continuously shifted toward the right, but the 10~--~20~keV X-ray amplitude dropped significantly compared to the 1.0 and 0.5~mm thick radiators. Therefore, for the 10~--~20~keV photons, the 1.0 and 0.5~mm thick radiators have the highest and second highest yields, respectively.   

\begin{figure}
	\begin{tabular}{cc}
		%\rput(0.5\width, 0.5\height){\psscalebox{20}{ssssss}}
		\begin{overpic}[width=4.1cm]{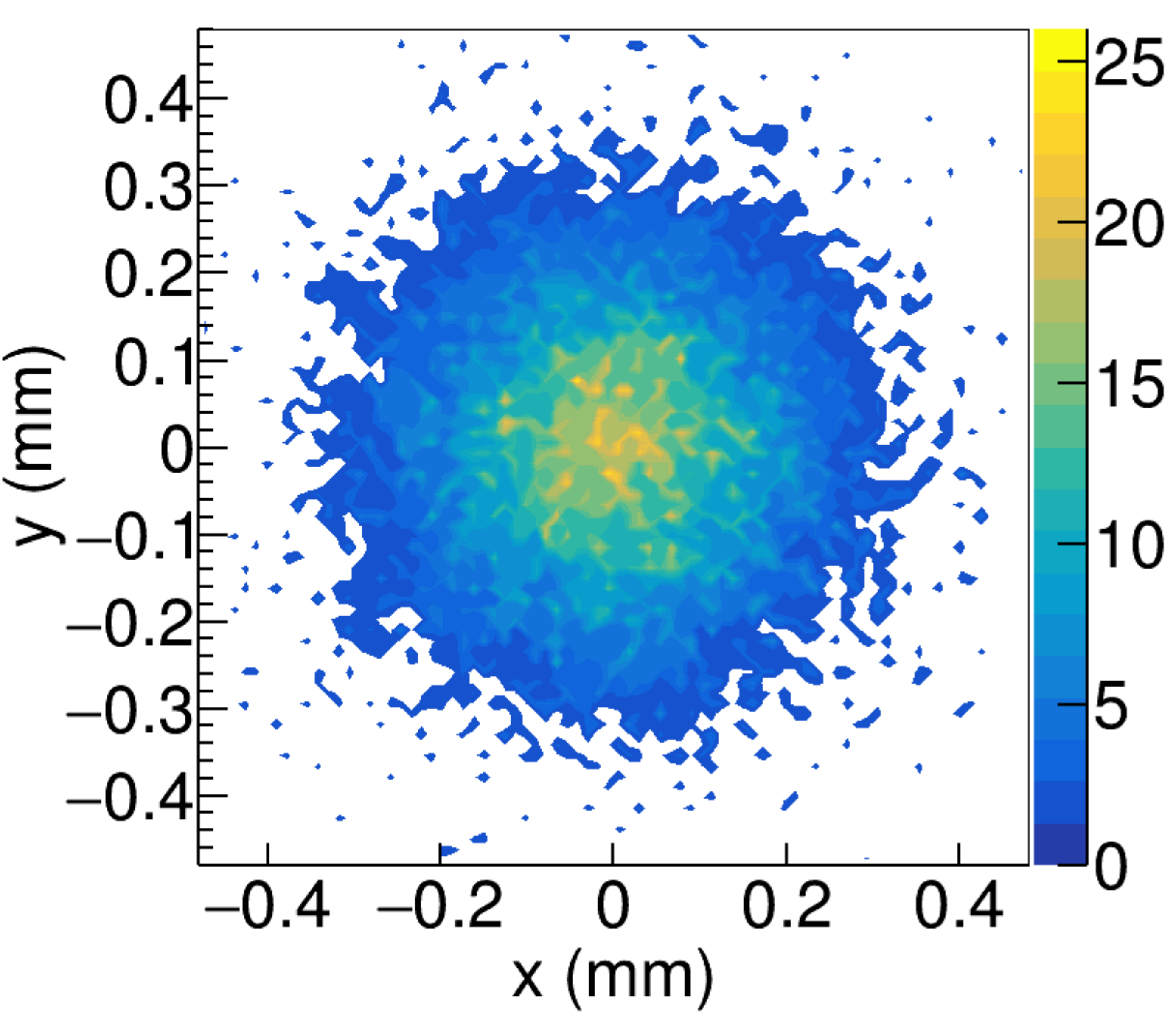} \put(77,77){(a)} \end{overpic}&
		\begin{overpic}[width=4.1cm]{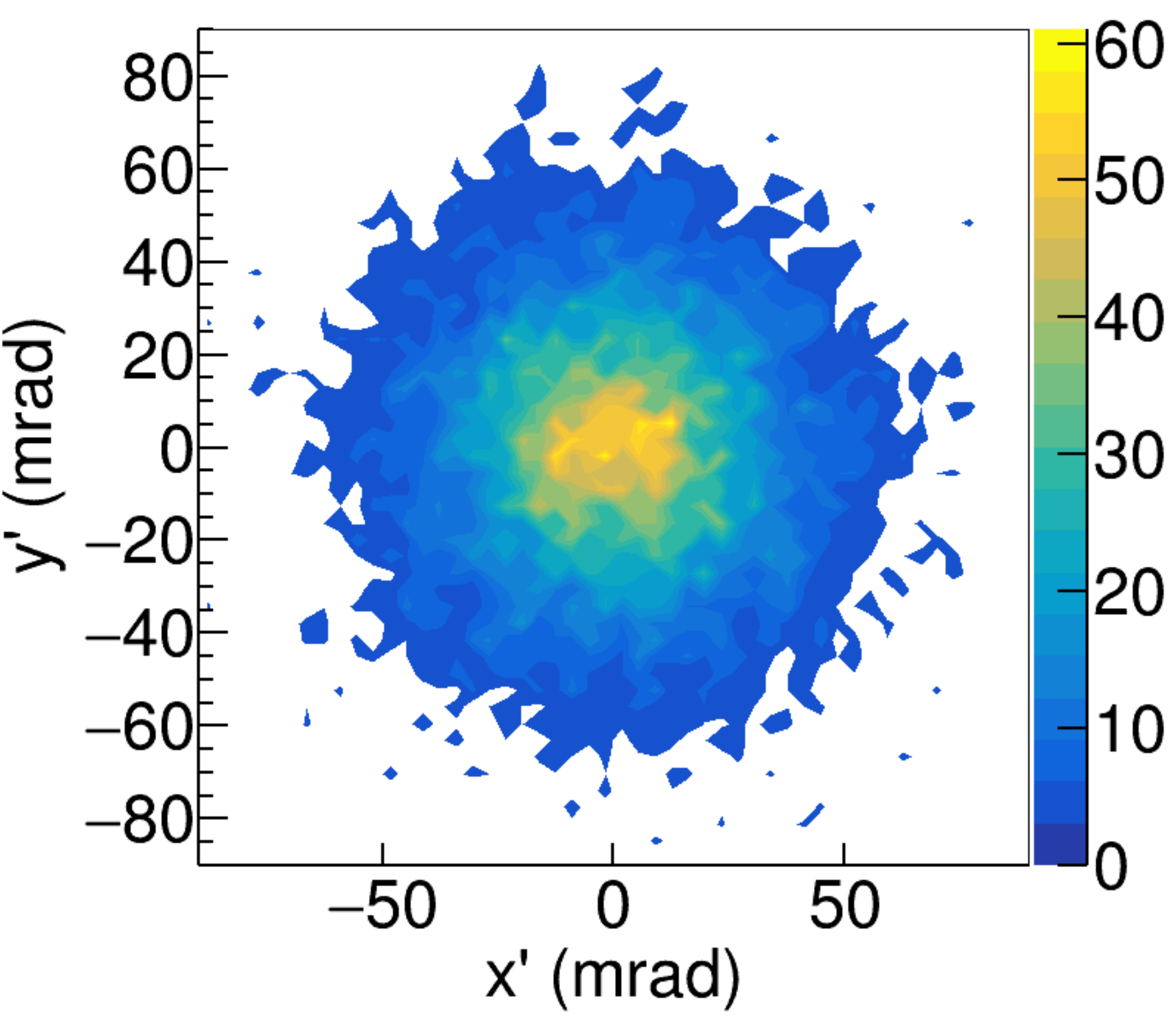} \put(77,77){(b)} \end{overpic}\\
		\begin{overpic}[width=4.1cm]{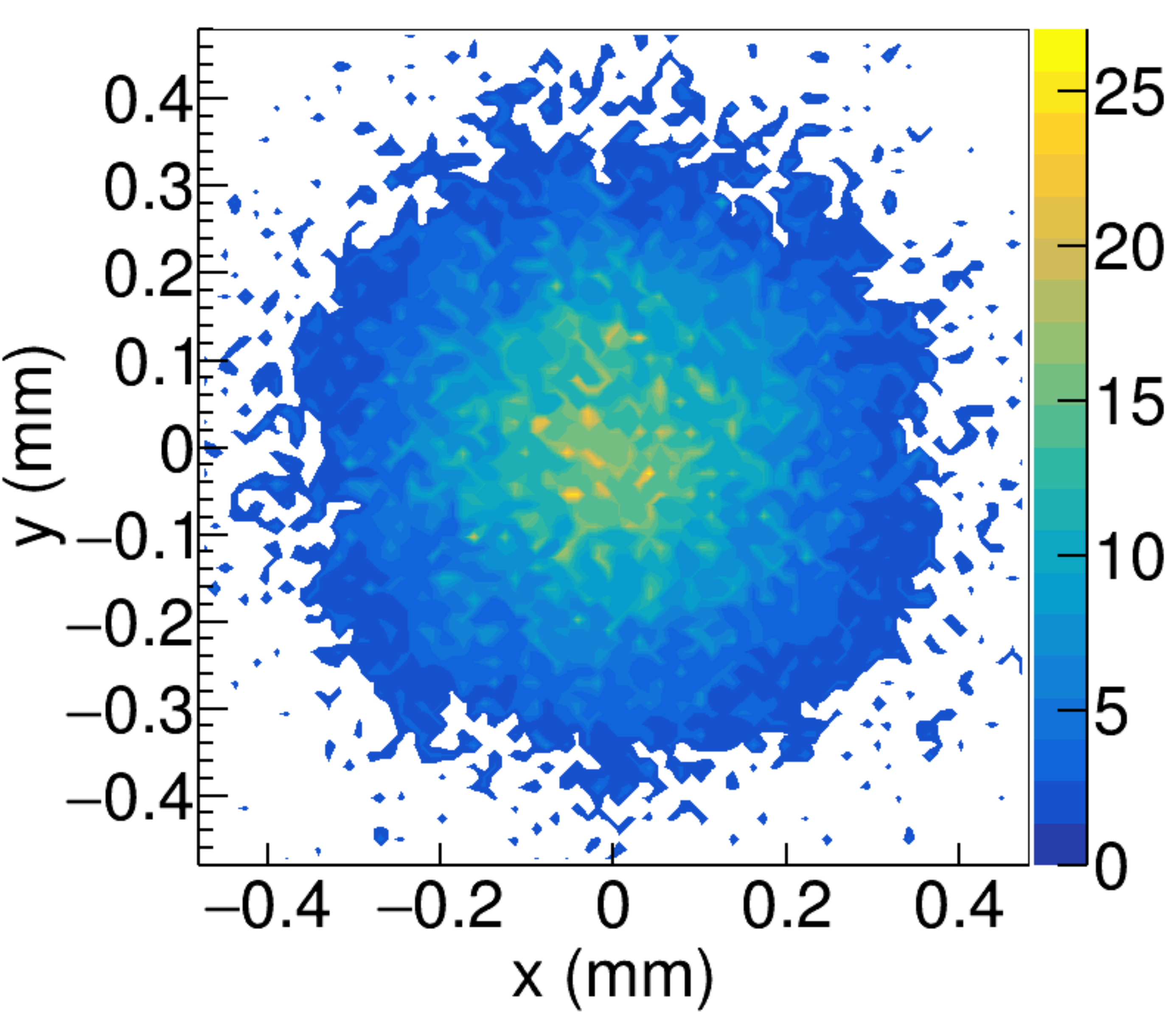} \put(77,77){(c)} \end{overpic}&
		\begin{overpic}[width=4.1cm]{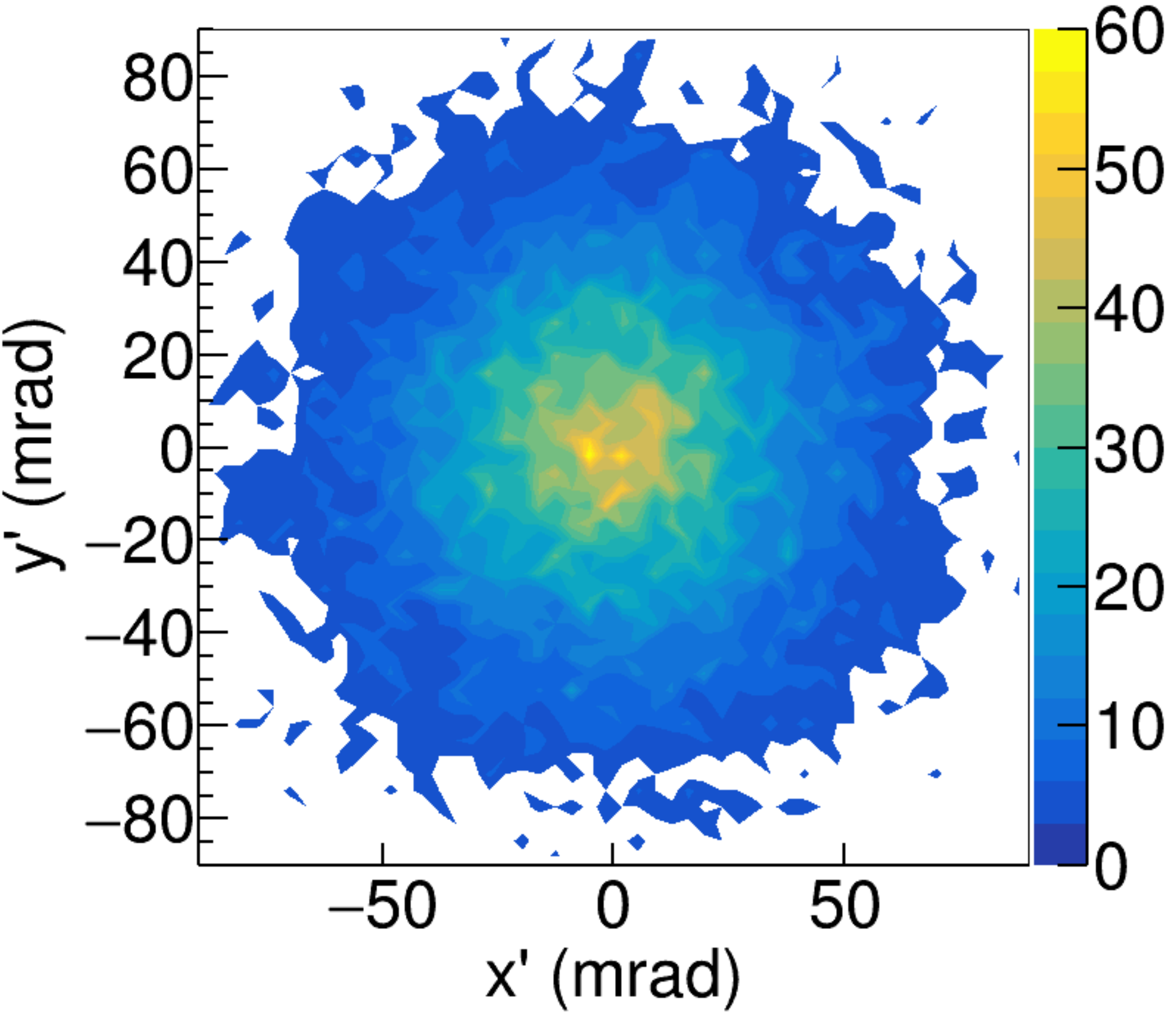} \put(77,77){(d)} \end{overpic}\\
	\end{tabular}
	\caption{Distribution of 10~--~20~keV X-rays at the detector. In (a) and (b) 0.5-mm-thick-foil was used; In (c) and (d) 1.0-mm-thick-foil was used. Sub-figures (a) and (c) are beam spatial distributions, while (b) and (d) are beam divergences.}
	\label{X-ray-dis-fig}
\end{figure}

The spatial and angular distributions of the X-rays from 1.0 and 0.5~mm thick radiators are given in Table.~\ref{X-ray-dis-tab} and in Fig.~\ref{X-ray-dis-fig}. As can be seen, the X-rays produced by the 0.5~mm radiator are fewer in number, but more compact (i.e. smaller angular and spatial distribution) and have higher intensity at their center. The thinner the radiator is, the fewer electrons are scattered, resulting in a more confined angular distribution. The compactness of the acceptance angle is important for X-ray optics, in terms of focusing/reflecting, and for the quality of the X-ray probe at the sample stage. Considering the limited space and budget, a 0.5~mm thick carbon radiator is the better option for our experimental setup with better beam quality. 
\begin{table*}
	\caption{Distribution of 10~--~20~keV X-rays generated when 1~pC electrons are incident on 0.5 and 1~mm thick carbon radiators.}
	\begin{ruledtabular}
		\begin{tabular}{lccc}
			{Parameter} & {Unit} & {0.5-mm-thick-foil} & {1-mm-thick-foil}\\
			 \hline
			number of photons	&						& 16468									& 20666	\\		
			RMS size $\sigma_{x}/\sigma_{y}$	&   mm  	& 0.240$\pm$0.001/0.257$\pm$0.001		& 0.260$\pm$0.001/0.280$\pm$0.001\\
			RMS divergence $\sigma_{x'}/\sigma_{y'}$&  mrad & 35.9$\pm$0.2/35.9$\pm$0.2 	& 41.1$\pm$0.2/41.1$\pm$0.2\\
		\end{tabular}
	\end{ruledtabular}
	\label{X-ray-dis-tab}
\end{table*}

\subsection{X-rays from Multifoil Radiator} 
\begin{figure}
	\includegraphics*[width=50mm]{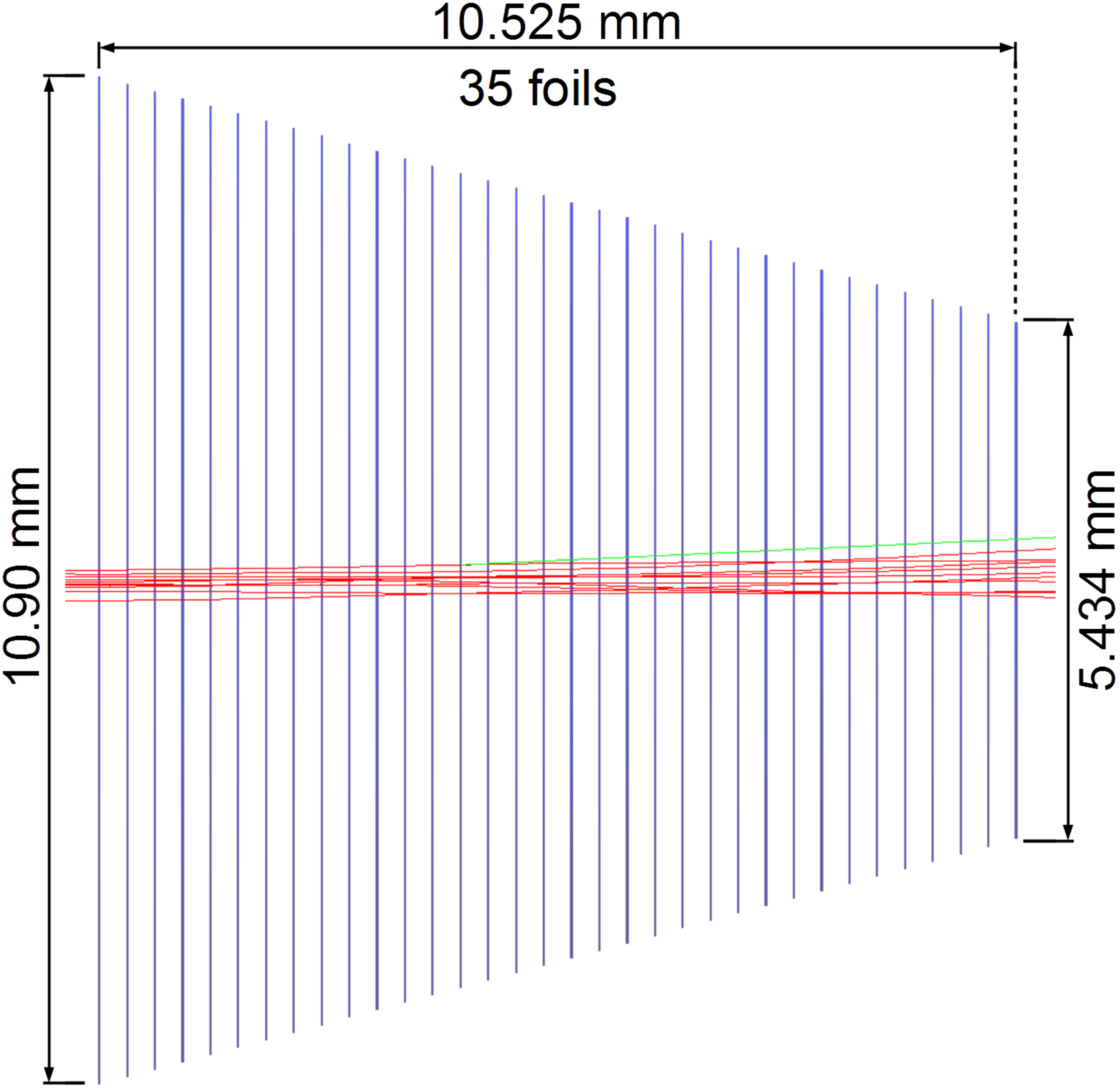}
	\caption{Geometry of the multifoil radiator (blue) and GEANT4 simulation of electrons traversing the multifoil radiator. The red and green lines represent the electron and photon trajectories, respectively.}
	\label{muti-foil-fig}
\end{figure}

\begin{figure}
	\begin{tabular}{cc}
	\begin{overpic}[width=4.1cm]{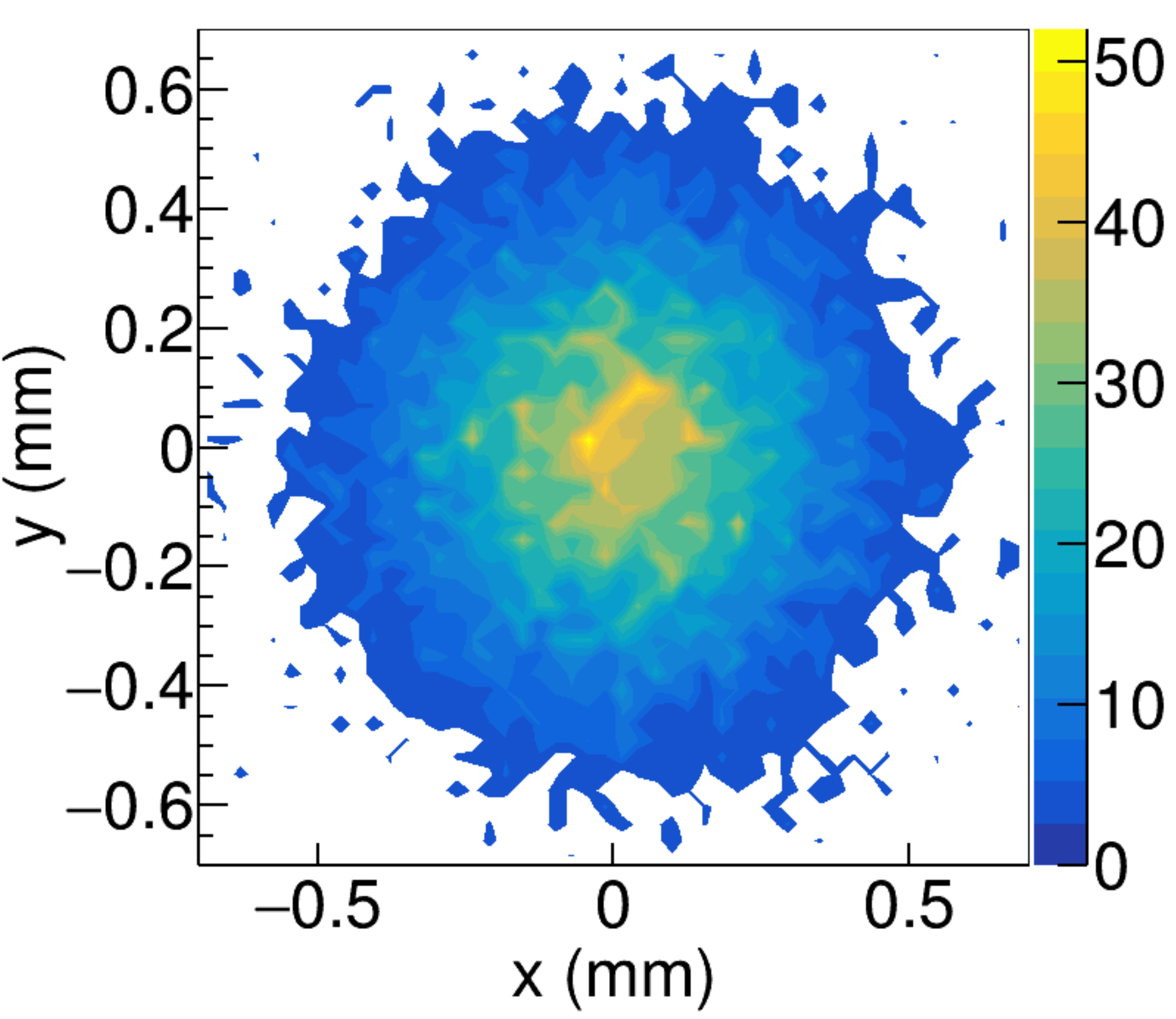} \put(77,77){(a)} \end{overpic}&
	\begin{overpic}[width=4.1cm]{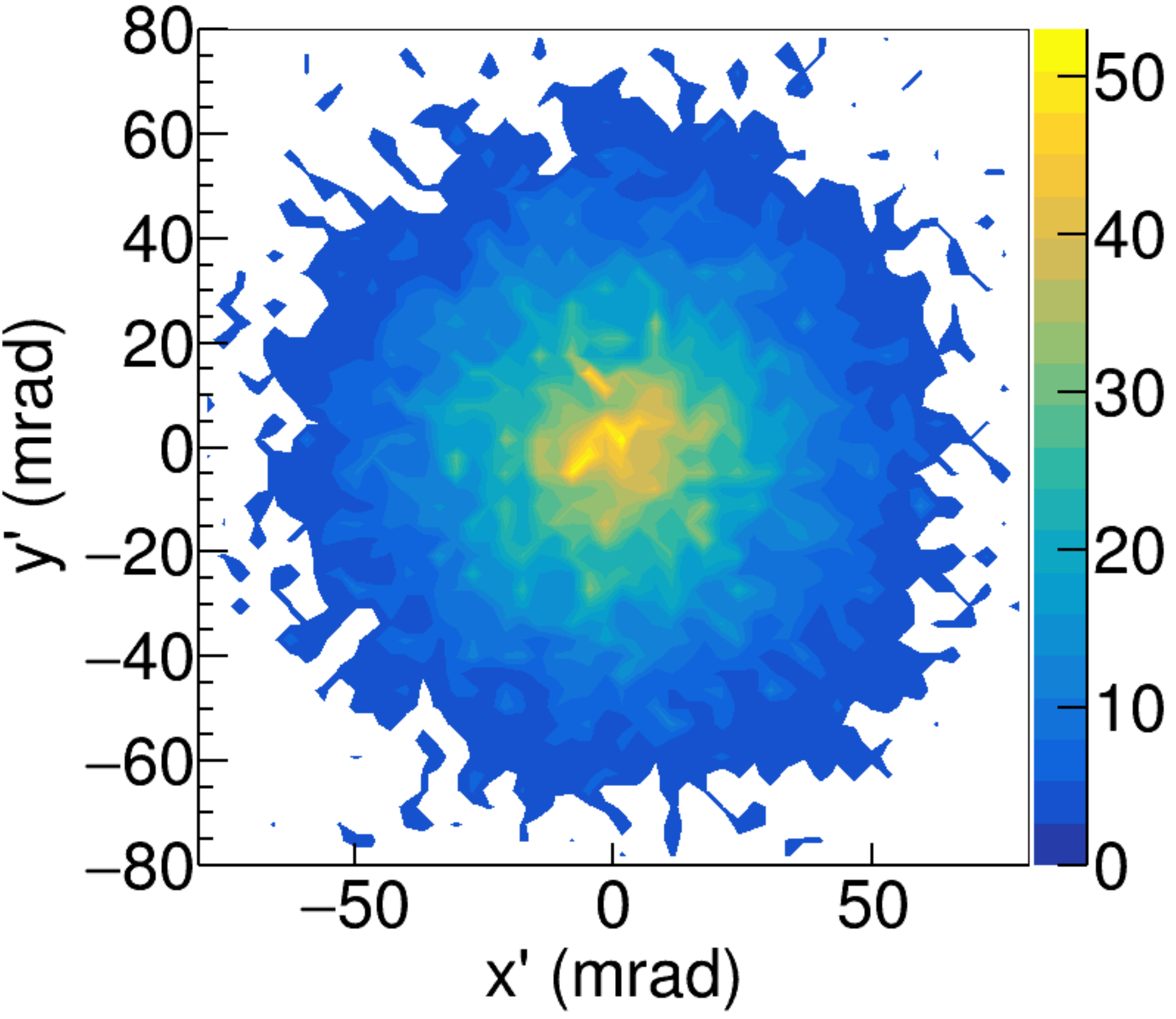} \put(77,77){(b)} \end{overpic}\\
	\end{tabular}
	\caption{Distribution of 10~--~20~keV X-ray at the end of the multifoil radiator: (a) beam size; (b) beam divergence.}
	\label{mfoil-X-ray-dis-fig}
\end{figure}

\begin{figure}
	\includegraphics*[width=80mm]{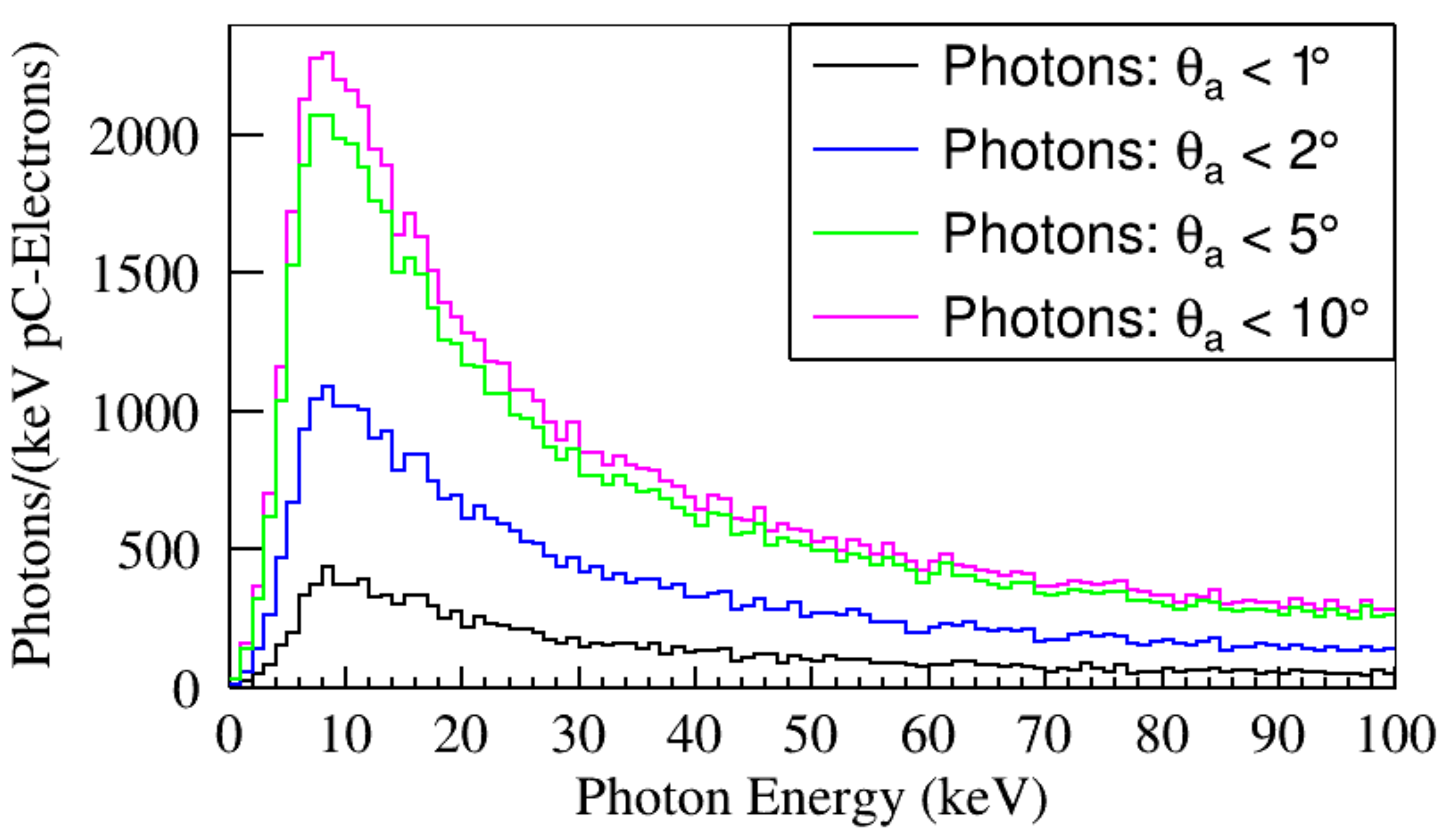}
	\caption{The X-ray energy spectrum generated by using multifoil radiator (1~pC incident electrons) with different cutoff polar angles.}
	\label{multifoil-Ene-fig}
\end{figure}

Based on the simulation results mentioned above, a graphite multifoil radiator will be used to generate high power THz pulses and X-rays. The multifoil radiator consists of 35 foil plates with successively decreasing radii stacked together as a truncated cone, as shown in Fig.~\ref{muti-foil-fig}. Each plate is 25~$\mu$m thick and placed with a 0.3~mm period. The diameter of the first/last plate is 10.90/5.434~mm and the cone angle is 30$^\circ$. The total thickness of the graphite foils is 0.875~mm, which provides the maximum 10~--~20~keV X-ray yield (0.5$\sqrt{2}$~$<$~0.875~$<$~1.0$\sqrt{2}$). 
The X-ray distributions from multifoil radiator, sampled immediately after the last foil, are summarized in Table.~\ref{mfoil-X-ray-dis-tab}; the beam-sizes and divergences are shown in Fig.~\ref{mfoil-X-ray-dis-fig}. The beam energy spectrum is shown in Fig.~\ref{multifoil-Ene-fig} for photons within 1$^\circ$, 2$^\circ$, 5$^\circ$, and 10$^\circ$ polar angles. The yield of 10~--~20~keV hard X-rays is approximately $10^3$ photons per 1~keV energy bin per picocoulomb of electrons. Our beamline can deliver bunches with 100~--~200~pC charges at a 100~Hz repetition rate. Thus, we can generate about $10^7$ hard X-rays per 1~keV energy bin per second or $10^5$ hard X-rays per 1~keV energy bin per pulse.
\begin{table}
	\caption{Distribution of 10~--~20~keV X-rays generated when 1~pC electrons are incident on carbon multifoil radiator.}
	\begin{ruledtabular}
		\begin{tabular}{lcc}
			{Parameter} & {Unit} & {Value} \\	
			 \hline
			number of photons	&	& 17869	\\		
			RMS size $\sigma_{x}/\sigma_{y}$	&   mm  &  0.280$\pm$0.001/0.285$\pm$0.002\\
			RMS divergence $\sigma_{x'}/\sigma_{y'}$	&  mrad  & 37.5$\pm$0.2/37.6$\pm$0.2\\
		\end{tabular}
	\end{ruledtabular}
	\label{mfoil-X-ray-dis-tab}
\end{table}

Owing to the finite exit aperture size, only a certain portion of the X-rays can be transmitted. We use the collection efficiency (CE) to describe the percentage of X-rays transmitted. For a given X-ray distribution, the CE is a function of the exit aperture size, which is described either by the radius or the acceptance angle $\theta_{a}$. The CE is the ratio of the number of X-rays transmitted ${ \textnormal{N}_{\textnormal{X-ray}} (\theta_{a}) }$ through the aperture subtended by $\theta_{a}$ and the total number of the X-rays generated ${N}_{\textnormal{X-ray,~tot}}$, i.e., 
%(${\sum\limits_{0 \leq \theta \leq \theta_{a}} \textnormal{N}_{\textnormal{X-ray}} (\theta) }$) and the total number of the X-rays (${N}_{\textnormal{X-ray,~tot}}$), i.e. 
\begin{equation}
\textnormal{CE} = \frac{\textnormal{N}_{\textnormal{X-ray}} (\theta_{a}) }{\textnormal{N}_{\textnormal{X-ray,~tot}}} \times 100\%.
\label{eq-col-ef}
\end{equation}

\begin{figure}
	\includegraphics*[width=70mm]{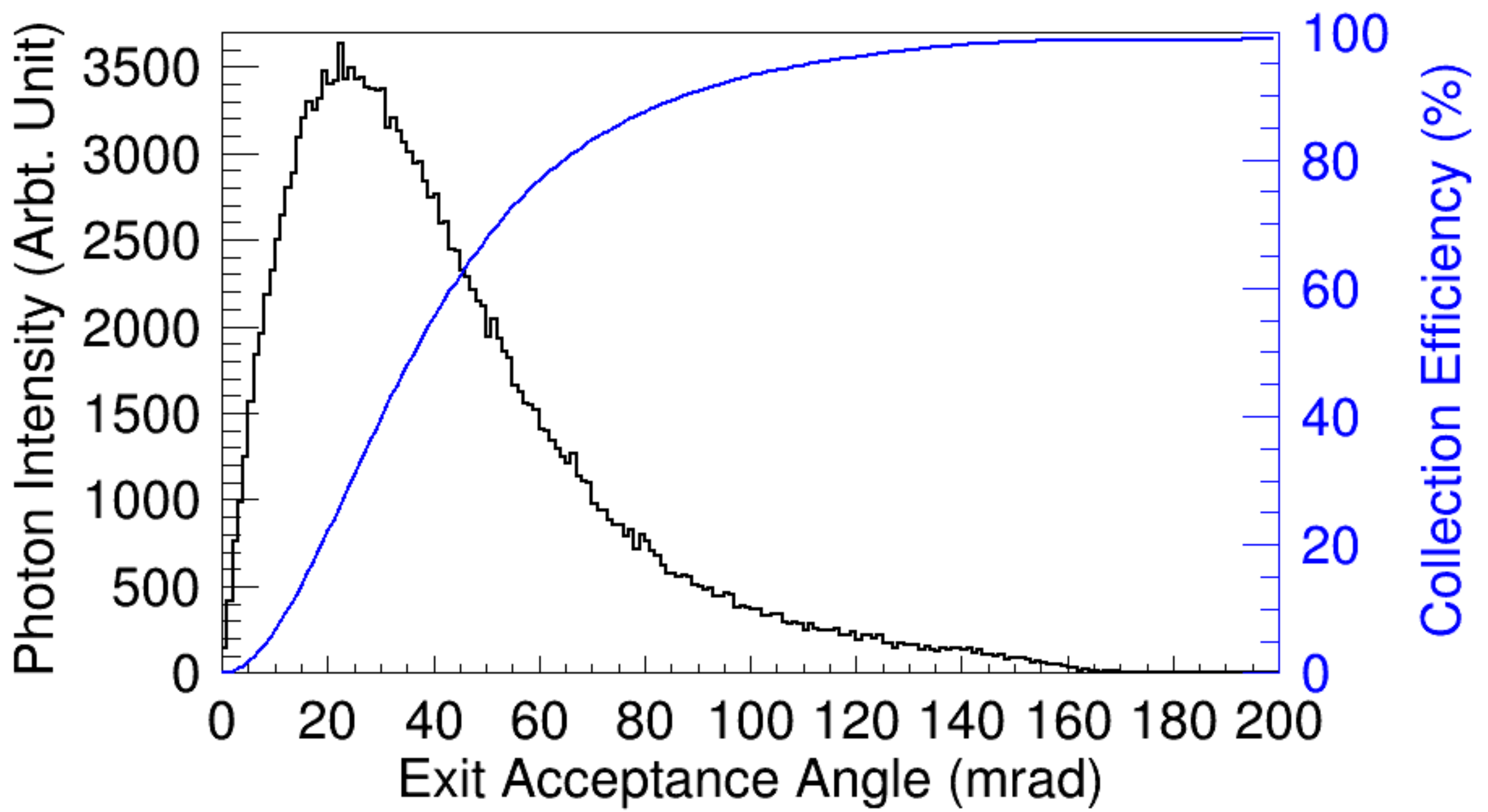}
	\caption{Scaling of the X-ray (generated by using the multifoil radiator) collection efficiency as function of the acceptance angle obtained by using GEANT4 simulation.}
	\label{col-eff}
\end{figure}

The CE of the X-rays from the multifoil radiators is shown in Fig.~\ref{col-eff}. The radius of the multifoil radiator X-ray exit is approximately 2~mm and the distance from the multifoil radiator to the exit is 57.7~mm, which subtends the acceptance angle of 34.8~mrad ($\sim$2.0$^\circ$). Hence, for our current setup, the CE of our multifoil radiator for 10~--~20~keV X-ray is around 43$\%$ with 2.0$^\circ$ beam divergence. 

\begin{figure}
	\includegraphics*[width=70mm]{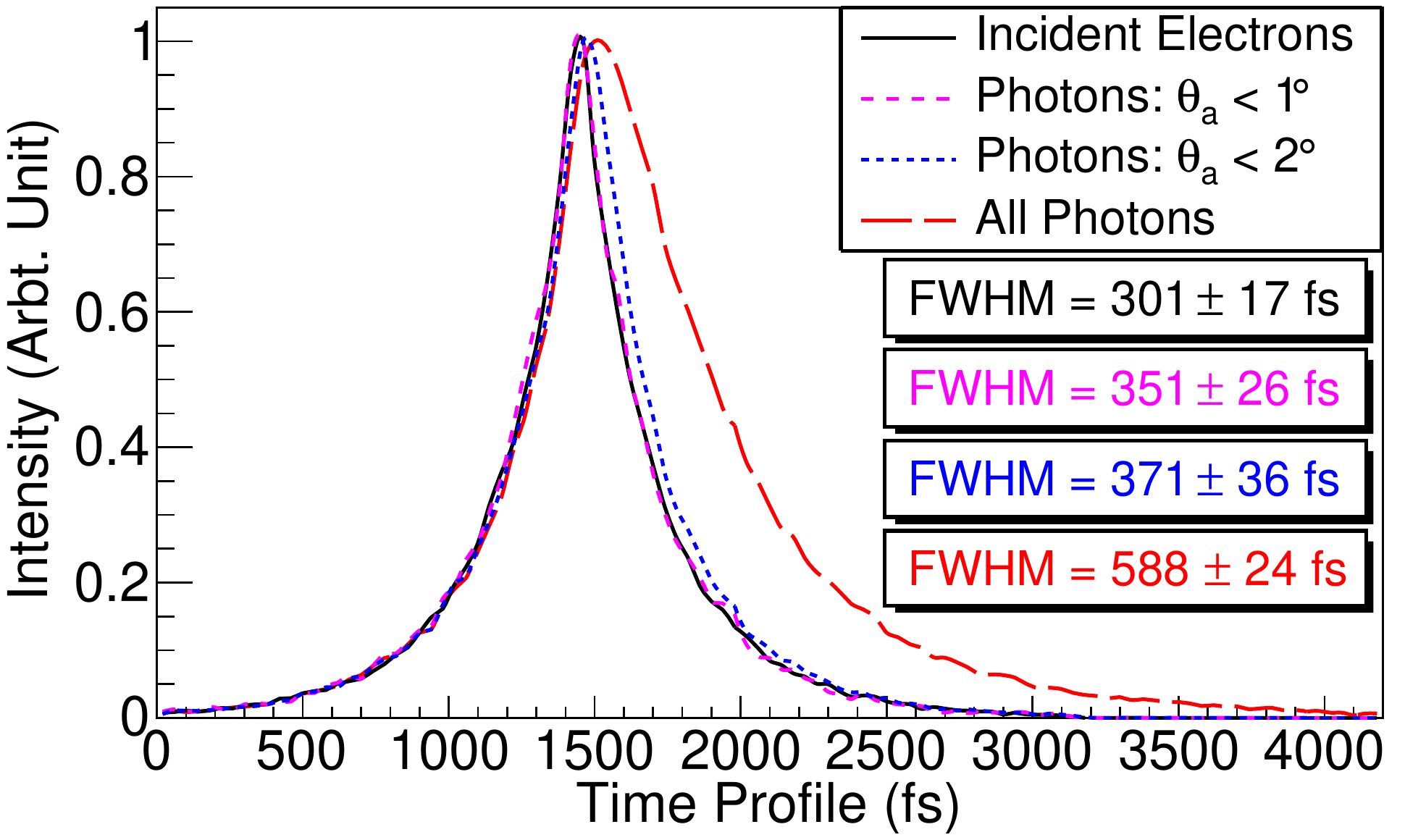}
	\caption{Time profile of the incident electron beam is shown in black. This electron beam resulted in the 10~--~20~keV X-ray photon time profile shown in red. The magenta/blue line is the time profile of the photons with acceptance angles smaller than 2$^\circ$/1$^\circ$.}
	\label{time-prof}
\end{figure}

\begin{figure}
	\includegraphics*[width=70mm]{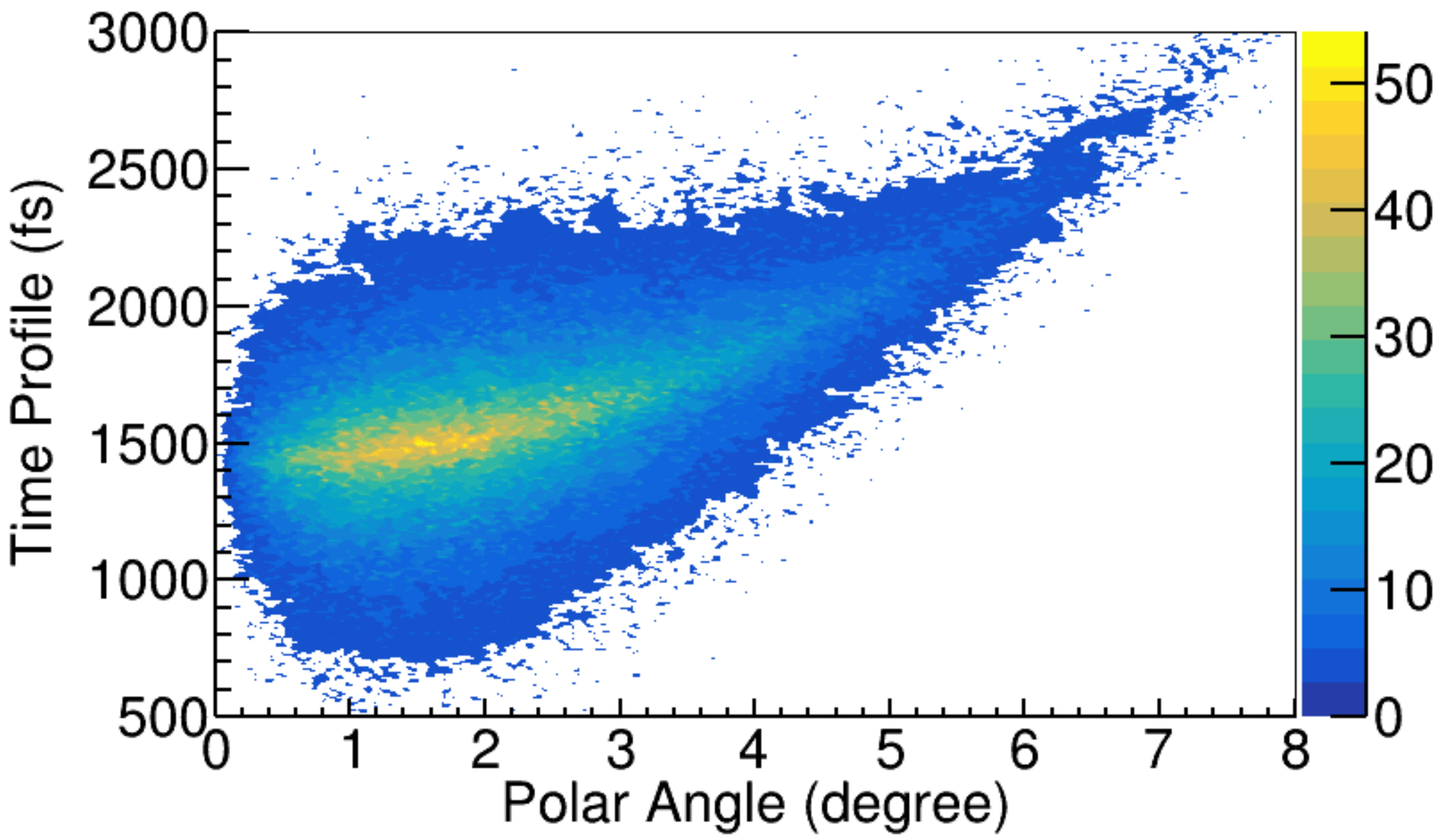}
	\caption{The arrival time of the X-rays vs. their polar angle. Photons with larger acceptance angles tend to have larger arrival time. %(Note that the arrival time on the vertical axis is obtained by subtracting 45 910~fs from the original arrival time.)
	}
	\label{time-vs-div}
\end{figure}

The time profiles of the X-ray pulses were investigated in the GEANT4 simulation. The generated electron time profile, shown in Fig.~\ref{time-prof} by the black curve, is similar to the one of $\phi=7.1^{\circ}$ in Fig~\ref{fig:RF_Jitter}. This electron beam incident on the multifoil radiator resulted in an X-ray pulse with 588$\pm$24~fs (FWHM) pulse durations, indicated by the red line. However, X-rays with acceptance angles smaller than 1$^\circ$ (magenta curve) and 2$^\circ$ (blue curve) have shorter pulse durations of 351$\pm$26~fs and 371$\pm$36~fs (FWHM), respectively. This is owing to the fact that the large angle photons pass through longer paths to arrive at the detector, and thus have larger arrival times and produce lengthened time profiles. It can be seen from the time profile vs. acceptance angle plot shown in Fig.~\ref{time-vs-div} that X-rays with higher polar angle tend to have larger arrival time. Therefore, one may shorten the X-ray time profile by choosing a smaller exit hole and cutting off large angle X-rays. 

\section{Radiator Heating}
The incident electron beam will heat up the radiator by depositing energy. Overheating the radiator may cause radiator damage and vacuum breakdown. We estimated the power deposited in a 25~$\mu$m thick graphite foil by using GEANT4 simulation. The beam parameters are the same as given in Table~\ref{sim-para}. A 200~pC bunch  deposits about 1.8~$\mu$J energy, which increases the foil temperature about 1.4~K at the center when electron beamsize is 0.1~mm. When operated at 100~Hz repetition rate with double pulse mode, the power deposited in the foil is 0.36~mW. Given the poor thermal conductivity between holder and foil, we assume most of the power is lost by thermal radiation. Assuming room temperature conditions (T=297~K), for a 2~mm diameter foil, the equilibrium temperature at the foil center is about 310~K when electron beamsize is 0.1~mm. Therefore, the temperature increase of radiator the foil caused by electron beam is insignificant. 

\section{Summary}
We described the detailed design of a novel high peak power THz pump source utilizes a multifoil radiator based on the idea of Ref.~\cite{Vinokurov}. We have extended it by adding X-ray probe. We have modified and improved the equations in Ref.~\cite{Vinokurov} to estimate THz peak power for a round shaped beams. ASTRA and ELEGANT simulations were used to transport and optimize the beam for maximum THz power. The results show that, when well optimized, a 200~pC bunch can generate peak THz power of 0.14~GW at the 7.1$^\circ$ RF phase, which is sufficient for most of the spectroscopy applications. The heating of the radiator by electrons is negligible because of the low beam current and energy deposit. 

The 10~--~20~keV X-ray yield optimized by scanning the thickness and material of different foils by using GEANT4 simulations. The results show that, for a 25~MeV electron beam, 0.5~--~1.0~mm thick carbon radiators have the highest yields of approximately $2\times10^3$ X-rays/(keV pC-electrons). Therefore, the total thickness of the carbon multifoil radiator is set to be 0.875~mm. The X-ray collection efficiency of the multifoil radiator holder is about 43$\%$ at 2$^\circ$ acceptance angle. Thus, we can transmit about $10^3$ hard X-rays/(keV pC-electrons) from our multifoil radiator. With a bunch charge of 200~pC and a 100~Hz repetition rate, we can transmit about $10^5$~X-rays/pulse and $10^7$~X-rays/s from the multifoil radiator.

The X-ray longitudinal time profile is around 400~--~600~fs depending on the acceptance angle. The broadening of the X-ray time profile due to the diverging effect of the beam is observed. An X-ray focusing mechanism near the source is under study to collimate the beam to prevent further broadening. 

\begin{acknowledgments}
	This work was supported by the World Class Institute (WCI) Program of the National Research Foundation of Korea (NRF) funded by the Ministry of Science, ICT and Future Planning (NRF Grant No. WCI 2011-001).
\end{acknowledgments}

%\bibliography{bibtex}
%\bibliographystyle{ieeetr}

%

\end{document}